\documentclass[pra,aps,showpacs,letterpaper,twocolumn,floatfix]{revtex4}

\usepackage{color}
\usepackage{graphicx}
\usepackage{amsmath}
\usepackage{bm}
\def\bd{\begin{displaymath}}
\def\be{\begin{equation}}
\def\ed{\end{displaymath}}
\def\ee{\end{equation}}

\def\vs{\textit{vs.}}
\def\vect#1{\bm{#1}}
\def\rvec{\vect r}
\def\kappavec{\vect{\kappa}}
\def\xivec{\vect{\xi}}

\def\abs#1{\lvert #1\rvert}
\def\bigabs#1{\bigl\lvert #1\bigr\rvert}
\def\avg#1{\left\langle #1\right\rangle}
\def\pd#1#2{\frac{\partial #1}{\partial #2}}

\def\eqref#1{Eq.~(\ref{#1})}% Use within sentence
\def\figref#1{Fig.~\ref{#1}}% Use within sentence
\def\secref#1{Sec.~\ref{#1}}% Use within sentence

\definecolor{jbcolor}{rgb}{0.6,0.0,0.1}

\definecolor{mycolor}{rgb}{0.5,0,0.5}

\begin{document}

\title{Exact Dynamics of Multicomponent Bose-Einstein Condensates in
        Optical Lattices in One, Two and Three Dimensions}
\author{R. Mark Bradley}
\affiliation{ Department of Physics,
        Colorado State University, Fort Collins, CO 80523, USA}
\author{James E. Bernard and L. D. Carr}
\affiliation{ Department of
        Physics, Colorado School of Mines, Golden, CO 80401, USA}

\date{\today}

\begin{abstract}

Numerous exact solutions to the nonlinear mean-field equations of
motion are constructed for multicomponent Bose-Einstein condensates
on one, two, and three dimensional optical lattices.  We find both
stationary and nonstationary solutions, which are given in closed
form.  Among these solutions are a vortex-anti-vortex array on the
square optical lattice and modes in which two or more components
slosh back and forth between neighboring potential wells.  We obtain
a variety of solutions for multicomponent condensates on the simple
cubic lattice, including a solution in which one condensate is at
rest and the other flows in a complex three-dimensional array of
intersecting vortex lines.  A number of physically important
solutions are stable for a range of parameter values, as we show by
direct numerical integration of the equations of motion.

\end{abstract}


\date{\today}

\maketitle

\section{Introduction}
\label{sec:introduction}

Two areas at the forefront of research in Bose-Einstein condensates
(BECs) over the last few years have been optical lattices and the
hyperfine degree of
freedom~\cite{leggett2001,lewenstein2006}. Optical lattices allow one
to explore, in both the mean-field and quantum regimes, the effects
of periodic potentials on bosons.  These studies complement the vast
body of knowledge concerning the behavior of fermions in periodic
potentials coming from solid state physics.  Optical lattices are
free of defects and disorder and the atomic potential has a simple
closed form.  In contrast, an electron in a solid is subject to a
complex, imperfectly known potential that is usually marred by
defects.

The hyperfine states of the atoms making up a BEC allow one to
construct exotic spin structures based on the occupation of different
hyperfine spin states of the form $|F,m_F\rangle$.  BECs of this
kind, which are called spinor or \emph{multicomponent}, have a vector
order parameter.  Research on multicomponent BECs has been
instrumental in reaching important milestones in BEC research,
including the creation of a quantum vortex and the subsequent
demonstration that a BEC made from a weakly interacting alkali gas is
superfluid~\cite{matthews1999,williams1999}.  Recently,
experimentalists have placed multicomponent BECs in optical
lattices~\cite{higbie2005,widera2005}.

In this article, we construct exact solutions to the mean-field
equations of motion for multicomponent BECs in one, two and three
dimensional optical lattices.  Band theory was invented as a tool to
analyze stationary solutions of the linear Schr\"odinger equation
with a periodic potential, a problem that does not have a solution in
closed analytic form for realistic potentials.  In contrast, for the
\emph{nonlinear} Schr\"odinger equation (NLS) with a sinusoidal
optical potential, exact, closed-form stationary solutions have been
discovered~\cite{carr2001b,carr2001c,carr2001d,deconinck2001,%
deconinck2002,hai2004,deconinck2003,bradleyRM2005}.
Here we generalize and extend previous work to an overarching and
rigorous treatment of certain classes of exact solutions describing
the dynamics of condensates with $s$ components in $D$ dimensions.
Our formalism permits us to construct exact, \emph{nonstationary}
solutions of the vector NLS.  This article brings together our new
work and past treatments into a single, general, rigorous analytical
framework.  At the same time, we elucidate the most experimentally
relevant and aesthetically pleasing solutions.  We also perform a
full nonlinear stability analysis where computationally tractable. A
number of surprising stability regimes present themselves, as we will
demonstrate.

The basic idea that leads to our exact solutions is to use the
nonlinearity to cancel the spatial variation of the potential,
leading to an effective free particle problem.  (Clearly, this is not
possible for the linear Schr\"odinger equation.)  This idea is due to
Bronski, Carr, Deconinck, and Kutz, who originally applied it to a
one-component, or scalar, BEC in a one-dimensional periodic
potential~\cite{carr2001b,carr2001c,carr2001d}.  Deconinck, Frigyik
and Kutz extended this work on one-component BECs to higher
dimensions~\cite{deconinck2001,deconinck2002}, and to multicomponent
condensates in one dimension~\cite{deconinck2003}, but not to
multicomponent BECs in two and three dimensions.  Although the
higher-dimensional Jacobi elliptic periodic potentials considered by
Deconinck~\textit{et al.} do not include the square, rectangular and
simple cubic optical potentials readily available in the lab, Hai and
coworkers have shown that the cancellation technique can be used to
construct solutions for a scalar condensate on a square optical
lattice~\cite{hai2004}.

Our extension of the cancellation technique to condensates with an
arbitrary number of components $s$ in a sinuoidal optical potential
of arbitrary dimension $D$ leads to an enormous number of new
solutions, many of great experimental import.  A crucial and
challenging step in the cancellation technique is to find a solution
ansatz that, for a given potential, allows the cancellation to occur.
At the same time, the solution ansatz must satisfy the free particle
Schr\"odinger equation.  A key aspect of our work is the introduction
of a novel, very general solution ansatz that allows the solution of
a wide range of problems.

Mean-field theory, which for a single-component condensate takes the
form of the scalar NLS or Gross-Pitaevskii
equation~\cite{dalfovo1999}, has been quite successful in describing
experiments on multicomponent BECs in optical
lattices~\cite{higbie2005,mur-petit2006}.  However, there are
important approximations underlying its use.  First, the tunneling or
hopping energy $t_h$ must be much larger than the on-site interaction
energy $U$ so that the system is far from the Mott insulating
regime~\cite{footnote1,footnote3}\nocite{rey2004}.  This means that
the potential barriers are not so high that the sites lose mutual
phase coherence, and that a full many body quantum Fock state
treatment is not needed~\cite{carr2007a}. Second, three-body and
other loss processes are neglected. Third, quantum fluctuations are
ignored, as is always the case when the NLS is applied to
BECs~\cite{dalfovo1999,carr2007a}. Fourth, any possible resonances
induced by the lattice or dimensional confinement are
neglected~\cite{olshanii1998}.  Fifth, when treating $D=1$ and $D=2$,
mean-field theory requires that the confining potential that reduces
the effective dimensionality have a length scale smaller than or of
the order of the healing length but larger than the scattering
length~\cite{carr2000e,petrov2000,petrov2000b}, i.e., the underlying
scattering process must remain three dimensional.

Three important conditions must be met if our cancellation technique
is to apply. First, we cannot include the low frequency harmonic trap
often, but not always, present in experiments.  Such a trap is used
to keep atoms from spilling off the edge of a finite lattice. We
require the potential to be sinusoidal, although we \emph{do} allow
the lattice constant to be different in each direction, leading to,
e.g., a rectangular lattice in two dimensions.  Second, for
condensates with three or more components, the mean-field theory
normally includes coherent couplings between different components of
the vector order parameter; for two components, such couplings are
prevented by angular momentum selection
rules~\cite{HoTL1998,ohmi1998}. We require incoherent couplings only,
which is appropriate for $s=2$.  For $s>2$ our treatment is always
correct for sufficiently short time scales~\cite{law1998}.  Our
treatment can also be correct at arbitrary times when the hyperfine
components are chosen from separate manifolds $F$.  Finally, we
cannot treat a mixture of scalar BEC's of different masses, despite
the fact that such a system can be described by a vector mean-field
theory with incoherent couplings only.

Although certain conditions must be satisfied for it to be
applicable, our cancellation technique yields a panoply of exact
solutions of great physical significance.  For example, for a
two-component condensate on a rectangular optical lattice, we find
temporally periodic solutions in which the optical lattice is divided
into two sublattices, and the condensate components oscillate back
and forth between these sublattices.  For the square optical lattice,
we find a vortex-anti-vortex array for a scalar condensate, while for
two-component condensates we obtain exotic solutions in which the
optical lattice is divided into a total of four sublattices, and the
condensate components move cyclically between these sublattices.  As
the dimension $D$ and the number of components $s$ are increased, the
number of solutions our technique generates grows rapidly.  The
number of solution types is so vast in three dimensions (3D) that,
for the sake of brevity, we limit our discussion to stationary
solutions with a high degree of symmetry and to two examples of
non-stationary solutions.

The article is organized as follows.  In Sec.~\ref{sec:principal}, we
introduce the mean-field equations of motion and the optical
potentials we will study.  In Sec.~\ref{sec:rigorous}, a complete and
rigorous treatment of exact dynamical solutions for arbitrary
dimensions $D$ and number of components $s$ is presented.  In
Sec.~\ref{sec:cases}, we treat select cases in detail, giving
examples of how to apply the results of Sec.~\ref{sec:rigorous}; of
particular interest are the vortex-anti-vortex array we find in 2D
and the array of intersecting vortex lines we find in 3D.  In
Sec.~\ref{sec:stability}, we present detailed stability studies of
important solution classes, including the vortex-anti-vortex array,
and an explicit connection to experimental units.  Finally, in
Sec.~\ref{sec:conclusion}, we conclude.

\section{The Mean-Field Equations of Motion}
\label{sec:principal}

Consider an $s$-component BEC in $D$ dimensions with incoherent
couplings between components.  The condensate is subject to an
optical potential formed by $D$ linearly polarized retroflected light
waves.  Let ${\bf k}_l$ be the wave vector of the $l^{\mathrm{th}}$
light wave, where $l\in\{1,\ldots , D\}$ will be called the
directional index.  We will restrict our attention to optical
lattices with ${\bf k}_l\cdot {\bf k}_{l'} = 0$ for $l\ne l'$. In
particular, we will study BECs in one-dimensional, square,
rectangular, and cubic optical lattices.  For a review of optical
potentials for neutral atoms, see, for instance,
Ref.~\cite{grimm2000}.

The mean-field equations of motion are
\begin{equation}
i\hbar{{\partial\psi_j}\over {\partial t}} =
\left[-{{\hbar^2}\over {2 m}}\nabla^2
	+ \left (\,\sum_{j'=1}^s g_{jj'}\vert\psi_{j'}\vert^2 \right)
	+ V_j\right]\psi_j\,,
\label{eqn:GP}
\end{equation}
where $\psi_j({\bf r},t)$ is the $j^{\mathrm{th}}$ component of the
condensate order parameter, $j\in\{1,\ldots,s\}$, and the position
${\bf r}$ is a vector in $D$ dimensions.  The $j^{\mathrm{th}}$
component of the order parameter may be written $\psi_j({\bf
        r},t)=\sqrt{n_j({ \bf r},t)}\exp[iS_j({ \bf r},t)]$, where
$n_j({ \bf r},t)$ is the number density of the $j^{\mathrm{th}}$
component at position ${\bf r}$ and ${ \bf v}_j({ \bf
        r},t)=(\hbar/m){\bm\nabla} S_j({ \bf r},t)$ is its velocity
at that point~\cite{dalfovo1999}.  Atoms in the $j^{\mathrm{th}}$
component are subject to the optical potential $V_j$.  The
coefficients $g_{jj'}$ of the nonlinear terms describe the binary
interaction of an atom in component $j$ and an atom in component
$j'$: explicitly, $g_{jj'}=4\pi\hbar^2 a_{jj'}/m$, where $a_{jj'}$ is
the $s$-wave scattering length and $m$ is the atomic mass, which, as
stated in Sec.~\ref{sec:introduction}, is assumed to be independent
of the component index $j$.  Note that the nonlinear coefficient
$g_{jj'}$ is renormalized by transverse confinement as briefly
alluded to in Sec.~\ref{sec:introduction} for $D=1$ and $D=2$; see
the references given there for more details.  The $n\times n$ real,
symmetric matrix $M\equiv \{g_{jj'}\}$ will be referred to as the
\emph{interaction matrix}.  We will assume that all of the diagonal
elements of $M$ are nonzero, as is the case in experiments.

The atoms of the $j^{\mathrm{th}}$ component are subject to the
optical potential
\begin{equation}
V_j({\bf r}) =
-{1\over 4} p_j \sum_{l=1}^{D} e_l^2 \cos^2({ \bf k}_l \cdot { \bf r}),
\label{eqn:potential}
\end{equation}
where $p_j$ is the atomic polarizability of an atom in the
$j^{\mathrm{th}}$ hyperfine state and $e_l$ is the electric field
amplitude of the $l^{\mathrm{th}}$ standing light wave.  The wave
vector ${\bf k}_l$ determines the lattice constant in the
$l^{\mathrm{th}}$ direction.  For convenience, we let $V_{jl}\equiv
{1\over 4} p_j e_l^2$.  Equation~(\ref{eqn:potential}) then becomes
\begin{equation}
V_j({\bf r}) = - \sum_{l=1}^{D} V_{jl} \cos^2 ({ \bf k}_l \cdot { \bf r}).
\label{eqn:potential2}
\end{equation}
We assume that all of the $p_j$'s and $e_l$'s are nonzero, so that
none of the $V_{jl}$'s vanish; if $V_{jl}$ were zero, the atoms of
the $j^{\mathrm{th}}$ component would be subject to an optical
potential independent of the $l^{\mathrm{th}}$ spatial coordinate.
Note that the atomic polarizability $p_j$ generally depends on the
component, or hyperfine state, $j$.

To illustrate how one arrives at the potential given by
Eq.~(\ref{eqn:potential}), consider the case $D = 2$. The total
electric field ${\bf E} = {\bf E}_1 + {\bf E}_2$, where
\begin{equation} {\bf E}_l = {\bf e}_l \cos({\bf k}_l\cdot{\bf
r}+\chi_l)\cos(c |{\bf k}_l| t +\phi_l)\,.\end{equation}
The vector ${\bf e}_l$ has constant, real components and is orthogonal
to ${\bf k}_l$, and $c$ is the speed of light in vacuum. By
shifting the location of the origin if necessary, one can arrange
for both of the spatial phases $\chi_l$ to vanish. The
temporally averaged intensity is then
\begin{eqnarray}
I &=&{1\over 2}e_1^2 \cos^2({\bf k}_1\cdot{\bf r}) + {1\over 2}e_2^2
\cos^2({\bf k}_2\cdot{\bf r})+\delta_{k_1,k_2}\nonumber\\
&&\times
{\bf e}_1\cdot{\bf e}_2\cos(\phi_1-\phi_2)\cos({\bf
k}_1\cdot{\bf r})\cos({\bf k}_2\cdot{\bf r})\,.
\end{eqnarray}
The $j^{\mathrm{th}}$ component of the BEC is subject to the external
optical potential $V_j = -{1\over 2} p_j I$.  The optical potentials
$V_j$ take the form~(\ref{eqn:potential}) with $D=2$ if and only if
the term in $I$ coming from the interference of the two light waves
vanishes.  If $k_1$ and $k_2$ differ, the interference term is zero
and the optical potential has rectangular symmetry. If $k_1 = k_2$,
on the other hand, the interference term vanishes if either ${\bf
        e}_1\cdot{\bf e}_2 = 0$ or $\cos(\phi_1-\phi_2) = 0$.  The
optical lattice is then simply a square lattice. Finally, if
$k_1=k_2$ and $0<\vert{\bf e}_1\cdot{\bf e}_2\cos(\phi_1-\phi_2)\vert
< e_1 e_2$, the structure of the optical lattice is more
complex~\cite{hemmerlich1992}.

The interference terms in the intensity can also be
made to vanish for $D = 3$: for example, we can choose the vectors
${\bf e}_j$ to be orthogonal to one another. If $k_1 = k_2 = k_3$,
the optical lattice has simple cubic symmetry.

For simplicity, in this paper we will confine ourselves to optical
potentials in which the interference terms vanish.  The potential is
then given by Eq.~(\ref{eqn:potential}).  It is worth noting, though,
that our solution techniques can be generalized to optical potentials
with nonzero interference terms.

\section{Methods of Constructing Exact Solutions to the
Mean-Field Equations of Motion}
\label{sec:rigorous}

The time evolution of the order parameter is described by the
mean-field equations of motion~(\ref{eqn:GP}) with the optical
potentials~(\ref{eqn:potential2}).  We will seek solutions to this
problem in which each of the effective potentials
\begin{equation}
U_j({\bf r},t) \equiv
V_j({\bf r}) + \sum_{j'=1}^s g_{jj'}\vert\psi_{j'}({ \bf r},t)\vert^2
\label{eqn:effective_potential}
\end{equation}
is constant.  Solutions of this kind will be referred to as
\emph{potential-canceling}~(PC) solutions and our solution technique
will be called the \emph{cancellation method} because for each
$j\in\{1,2,\ldots,s\}$, the spatial variation of the optical
potential $V_j({\bf r})$ is canceled by the variation of the term
$\sum_{j'=1}^s g_{jj'}\vert\psi_{j'}({\bf r},t)\vert^2$, rendering
the effective potential $U_j$ constant.  PC solutions reduce the
coupled nonlinear mean-field equations of motion~(\ref{eqn:GP}) to
uncoupled linear Schr\"odinger equations with constant potential:
\begin{equation}
i\hbar{{\partial\psi_j}\over {\partial t}} =
\left(-{{\hbar^2}\over {2 m}}\nabla^2 + U_j\right)\psi_j\,,
\label{eqn:free}
\end{equation}
for $j\in \{1,2,\ldots s\}$.

For the $U_j$'s to be constant, the $\psi_j$'s must be linear
combinations of terms that vary sinusoidally with position.  To be
precise, we seek solutions of the form
\begin{equation}
\psi_j({ \bf r},t) =
e^{-i\Omega_j t}\sum_{l=0}^{D}
A_{jl}\cos({\bf k}_l\cdot{\bf r}) e^{-i\omega_l t},
\label{eqn:solution}
\end{equation}
where $\hbar\omega_l =\hbar^2 k_l^2 / 2m$ is the energy of a free
particle of mass $m$ with wave number $k_l$ in the absence of
nonlinearity.  Note that the possible values of the directional index
$l$ have been extended to include $l=0$ in order to simplify the
notation: ${\bf k}_0\equiv 0$ so that the $l=0$ term gives rise to a
constant offset in the order parameter.  The coefficients $A_{jl}$ in
our solution ansatz~(\ref{eqn:solution}) are in general complex,
while the frequencies $\Omega_j$ are real.  The $A_{jl}$'s are
constrained by the requirement that the effective potentials $U_1,
U_2,\ldots ,U_s$ are constant.  These constraints will be discussed
in detail below.

Substituting Eq.~(\ref{eqn:solution}) into Eq.~(\ref{eqn:GP}), we see
that
\begin{equation}
V_j = - \sum_{j'=1}^s  g_{jj'}\vert\psi_{j'}\vert^2 + \hbar\Omega_j ,
\label{eqn:potentialPsi}
\end{equation}
for $j\in \{1,2,\ldots s\}$. This means that the effective potential
$U_j$ takes on the constant value $\hbar\Omega_j$.  Inserting
Eq.~(\ref{eqn:solution}) into Eq.~(\ref{eqn:potentialPsi}) and
comparing the resulting expression for $V_j$ with
Eq.~(\ref{eqn:potential2}), we obtain
\begin{eqnarray}
\sum_{l=1}^{D} V_{jl} \cos^2 ({ \bf k}_l \cdot { \bf r})
= \sum_{l=0}^D \sum_{l'=0}^D
\left( \sum_{j'=1}^s g_{jj'}A_{j'l}A^\ast_{j'l'}\right)
\nonumber\\
\times\cos ({ \bf k}_l \cdot { \bf r})
\cos ({ \bf k}_{l'} \cdot { \bf r}) e^{-i(\omega_l-\omega_{l'})t}
-\hbar \Omega_j,
\label{eqn:comparison}
\end{eqnarray}
for $j\in\{1,2,\ldots,s\}$.  Equation~(\ref{eqn:comparison}) yields a
set of algebraic equations that the coefficients $A_{jl}$ must
satisfy.  The first set of equations ensure that the cross terms on
the right hand side of Eq.~(\ref{eqn:comparison}) vanish.
Specifically, for each pair of integers $(l, l')$ with $0\le l < l'
\le D$, one obtains a set of conditions.  If $\omega_l \ne
\omega_{l'}$, we must have
\begin{equation}
\sum_{j'=1}^s g_{jj'}A_{j'l}A^\ast_{j'l'} = 0
\label{eqn:conditionNE}
\end{equation}
for $j\in\{1,\ldots, s\}$.  On the other hand, if $\omega_l =
\omega_{l'}$, it is sufficient to impose the weaker conditions
\begin{equation}
\Re\left(\sum_{j'=1}^s
 g_{jj'}A_{j'l}A^\ast_{j'l'}\right) = 0\,.
\label{eqn:conditionE}
\end{equation}
Equating the coefficients of the terms that are proportional to
$\cos^2({ \bf k}_l \cdot { \bf r})$ on either side of
Eq.~(\ref{eqn:comparison}), we see that
\begin{equation}
V_{jl} = \sum_{j'=1}^s g_{jj'} \vert A_{j'l}\vert^2
\label{eqn:VA}
\end{equation}
for $j\in\{1,2,\ldots,s\}$ and $l\in\{1,2,\ldots,D\}$.  (Note that
this equation does not apply for $l=0$.)  Finally, the constant term
on the right hand side of Eq.~(\ref{eqn:comparison}) must vanish, and
so
\begin{equation}
\hbar\Omega_{j} = \sum_{j'=1}^s g_{jj'} \vert A_{j'0}\vert^2
\label{eqn:OmegaA}
\end{equation}
for $j\in\{1,2,\ldots,s\}$.

The task of finding solutions to the coupled nonlinear partial
differential equations~(\ref{eqn:GP}) has now been reduced to solving
a system of algebraic equations: Eq.~(\ref{eqn:VA}) must be solved
for the coefficients $A_{jl}$ subject to the
conditions~(\ref{eqn:conditionNE}) or (\ref{eqn:conditionE}) for each
pair $(l, l')$ with $l < l'$. A solution to these equations does not
necessarily exist.  If a solution does exist, Eq.~(\ref{eqn:OmegaA})
yields the frequencies $\Omega_{j}$.

The solution, if it exists, is not uniquely specified by the system
of algebraic equations.  To see this, let
\begin{equation}
{\bf A}_l \equiv (A_{1l}, \ldots, A_{sl})^T
\end{equation}
for $l\in\{0,1,2,\ldots,D\}$.  Equation~(\ref{eqn:VA}) determines the
norm of the vector ${\bf A}_l$ for $l\in\{1,2,\ldots,D\}$ but does
not constrain the magnitude of ${\bf A}_0$.  In addition,
Eqs.~(\ref{eqn:conditionNE}) and (\ref{eqn:conditionE}) with $l=0$
place constraints on the direction of ${\bf A}_0$ but not its norm.
The quantity $|{\bf A}_0|^2$ is therefore a free parameter.

The length of the vector ${\bf A}_0$ is determined if the spatial
average of the total density is given, as we will now establish.  At
time $t$, the number density of the $j^{\mathrm{th}}$ component at
position ${\bf r}$ is $n_j({\bf r}, t) \equiv |\psi_j({\bf r},
t)|^2$.  Let $\langle f \rangle$ denote the spatial average of an
arbitrary function $f({\bf r})$.  Using Eq.~(\ref{eqn:solution}), we
find that the spatial average of the total number density
\begin{equation}
\langle n \rangle \equiv \sum_{j=1}^s \langle n_j \rangle
\end{equation}
is given by
\begin{equation}
\langle n \rangle = |{\bf A}_0|^2 + {1\over 2}\sum_{l=1}^D |{\bf A}_l|^2.
\label{eqn:A_0}
\end{equation}
Equation~(\ref{eqn:A_0}) has a solution for $|{\bf A}_0|^2$ if and
only if
\begin{equation}
2\langle n \rangle\ge \sum_{l=1}^D |{\bf A}_l|^2\,;
\end{equation}
if this condition is met, $|{\bf A}_0|^2$ is uniquely specified.

Equation~(\ref{eqn:conditionNE}) states that the vector
$(A_{1l}A^\ast_{1l'},\ldots, A_{nl}A^\ast_{nl'})^T$ is in the kernel
of $M$, while if Eq.~(\ref{eqn:conditionE}) applies, the real part of
this vector must be in the kernel of $M$.  For this reason, most (but
not all) of the solutions we obtain will be for cases in which the
atomic interactions are such that $\det M = 0$.

We will now consider three particularly interesting and physically
important special cases in which the algebraic conditions that must
be solved to yield a solution simplify dramatically.  These special
cases will be referred to as Special Cases A, B and C.  We will also
provide an example that shows that the formalism just developed
yields solutions to the mean-field equations of motion even when the
interaction matrix is nonsingular.  This example appears in
Subsection~\ref{ssec:nonsingular}.

\subsection{Factorizable Equations of Motion}
\label{ssec:factorizable}

A particularly simple special case is obtained when the rank of $M$
is unity.  This is true to an excellent approximation for the
two-component condensates first produced by the JILA group that
consist of two different hyperfine spin states of $^{87}$Rb:
$g_{11}$, $g_{12}$ and $g_{22}$ are known to the $1\%$ level, and are
in the proportion 1.03 : 1 : 0.97 \cite{hall1998}.  As a result,
$\det M/\mathop{\mathrm{Tr}} M$ is zero to within experimental error.

Because $M$ has rank 1 and is a symmetric matrix, there are nonzero,
dimensionless, real numbers $\lambda_{j}$ and a $\sigma=\pm 1$ such
that
\begin{equation}
g_{jj'} = \sigma g\lambda_{j} \lambda_{j'}
\label{eqn:factor}
\end{equation}
for all $j$ and $j'$.  The quantity $g$ is a positive constant with
dimensions of energy times volume, which is inserted in
Eq.~(\ref{eqn:factor}) to render the $\lambda_j$'s dimensionless.
The magnitude of $g$ is arbitrary but fixed and, if desired, may be
taken to be the typical magnitude of the interaction coefficients
$g_{jj'}$.

Let ${\bf L} \equiv (\lambda_1, \lambda_2, \dots,\lambda_s)^T$.
Equation~(\ref{eqn:factor}) may then be written
\begin{equation}
M = \sigma g{\bf L}{\bf L}^T,
\label{eqn:factor2}
\end{equation}
showing that when the rank of the interaction matrix is 1, $M$ can be
factored.

Let $\Lambda$ be the $s\times s$ matrix with elements
\begin{equation}
\Lambda_{jj'} \equiv \lambda_{j}\delta_{j,j'}\,.
\end{equation}
For each pair of directional indices
$(l,l')$ with $0\le l < l' \le D$, Eq.~(\ref{eqn:conditionNE})
reduces to the single condition
\begin{equation}
{\bf A}_{l}^\dagger \Lambda {\bf A}_{l'} = 0\,,
\label{eqn:factCondNe}
\end{equation}
while Eq.~(\ref{eqn:conditionE}) becomes
\begin{equation}
\Re({\bf A}_{l}^\dagger \Lambda {\bf A}_{l'}) = 0\,.
\label{eqn:factCondE}
\end{equation}
The relations~(\ref{eqn:VA}) are now
\begin{equation}
{1\over 4} p_j e_l^2
= \sigma g\lambda_j {\bf A}^\dagger_l\Lambda {\bf A}_l\,,
\label{eqn:factVA}
\end{equation} where $j \in\{1,\ldots,s\}$
and $l\in\{1,2,\ldots,D\}$.  Equation~(\ref{eqn:factVA}) has a
solution if and only if
\begin{equation}
p_j = p \lambda_j,
\label{eqn:polar}
\end{equation}
for all $j$, where $p$ is a nonzero real constant.  If this is the
case, then
\begin{equation}
V_j = \lambda_j V,
\label{eqn:factPotential}
\end{equation}
where
\begin{equation}
V \equiv -{1\over 2} p I =
-{1\over 4} p\sum_{l=1}^ D e_l^2 \cos^2 ({\bf k}_l\cdot {\bf r})\,.
\label{eqn:defnV}
\end{equation}
Equation~(\ref{eqn:factVA}) then becomes
\begin{equation}
\sigma g{\bf A}^\dagger_l\Lambda {\bf A}_l = {1\over 4} p e_l^2\,;
\label{eqn:VAl}
\end{equation}
this holds for $l\in{1,2,\ldots,D}$. Equation~(\ref{eqn:OmegaA})
shows that
\begin{equation}
\Omega_j = \lambda_j\Omega,
\label{eqn:OmegajOmega}
\end{equation}
where
\begin{equation}
\Omega \equiv {\sigma g\over\hbar} {\bf A}^\dagger_0\Lambda {\bf A}_0.
\label{eqn:OmegaA0}
\end{equation}
Finally, Eq.~(\ref{eqn:potentialPsi}) reduces to
\begin{equation}
V = -\sigma g\bm{\psi}^\dagger\Lambda\bm{\psi} + \hbar\Omega,
\label{eqn:potentialPsi2}
\end{equation}
where
\begin{equation}
\bm{\psi}\equiv (\psi_1,\ldots, \psi_s)^T
\end{equation}
is the vector order parameter.

As before, $|{\bf A}_0|^2$ is a free parameter unless an additional
constraint is applied.  If the spatial average of the total density
$\langle n \rangle$ is given and
\begin{equation}
\langle n \rangle \ge {1\over 2}\sum_{l=1}^D |{\bf A}_l|^2\,,
\end{equation}
then
\begin{equation}
|{\bf A}_0|^2 = \langle n \rangle - {1\over 2}\sum_{l=1}^D |{\bf A}_l|^2.
\label{eqn:lengthA0}
\end{equation}

If $M$ has rank 1 and the condition~(\ref{eqn:polar}) holds, we call
the mean-field equations of motion \emph{factorizable}.  For this
case, which we will refer to as \emph{Special Case A}, the equations
of motion~(\ref{eqn:GP}) assume the simpler form
\begin{equation}
i\hbar{{\partial\bm{\psi}}\over {\partial t}} =
-{{\hbar^2}\over {2m}}\nabla^2 \bm{\psi} +\sigma g
(\bm{\psi}^\dagger\Lambda\bm{\psi})\Lambda\bm{\psi} +
V\Lambda\bm{\psi}\, .
\label{eqn:separableEom}
\end{equation}
To find exact solutions to the coupled nonlinear partial differential
equations~(\ref{eqn:separableEom}), Eqs.~(\ref{eqn:VAl}) and
(\ref{eqn:lengthA0}) must be solved for ${\bf A}_0, {\bf A}_1,\ldots,
{\bf A}_D$ subject to the conditions~(\ref{eqn:factCondNe}) or
(\ref{eqn:factCondE}) for each pair $(l, l')$ with $0\le l < l'\le
D$. If a solution to these equations exists,
Eqs.~(\ref{eqn:OmegajOmega}) and (\ref{eqn:OmegaA0}) yield the
frequencies $\Omega_{j}$.

In Appendix~\ref{app:lambda} we demonstrate that if the mean-field
equations of motion are factorizable, an additional simplification
can be made: without loss of generality, all of the $\lambda_j$'s may
be taken to be of unit modulus.  Therefore, for the remainder of the
paper, when we discuss Special Case A, we will assume that each of
the $\lambda_j$'s is equal to $\pm 1$.

\subsubsection{Constructing Exact Non-stationary Solutions Using a
Transformation}
\label{sssec:P-set}

If we have a PC solution $\bm{\psi}$ to the equation of
motion~(\ref{eqn:separableEom}), under certain circumstances we can
construct new solutions by transforming $\bm{\psi}$.  Let $P$ be an
invertible $s\times s$ matrix, and suppose that $\bm{\psi}$ is given
by Eq.~(\ref{eqn:solution}) and satisfies
Eq.~(\ref{eqn:separableEom}).  We set
\begin{equation}
\bm{\zeta}({\bf r},t) =  T(P)\bm{\psi}({\bf r},t),
\label{eqn:transformation}
\end{equation}
where
\begin{equation}
T(P)\equiv\exp(-i\Omega\Lambda t) P \exp(i\Omega\Lambda t).
\label{eqn:transformation2}
\end{equation}
From Eq.~(\ref{eqn:solution}), it follows that if
\begin{equation}
P^\dagger\Lambda P = \Lambda,
\label{eqn:transfCond}
\end{equation}
then $\bm{\zeta}({\bf r},t)$ is also a solution to the equation of
motion~(\ref{eqn:separableEom}). An important aspect of this
transformation is that even if $\bm{\psi}({\bf r},t)$ is stationary,
$\bm{\zeta}({\bf r},t)$ can turn out to be nonstationary.  Thus, by
transforming a single solution $\bm{\psi}$, we obtain a set of
stationary and nonstationary solutions.  We will call each such set a
\emph{P-set}.

For a given $\Lambda$, let $S(\Lambda)$ be the set of invertible
$s\times s$ matrices $P$ that satisfy Eq.~(\ref{eqn:transfCond}).  It
is straightforward to show that the set of matrices $T(P)$ with $P\in
S(\Lambda)$ forms a group.  We will call this group by $G(\Lambda)$.
Later in the paper we will study two examples in which $G(\Lambda)$
is a continuous group, i.e., it has an uncountably infinite number of
elements.  As a result, the $P$-set is uncountably infinite in these
examples.

\subsection{Factorizable Equations of Motion with Equal Atomic
Polarizabilities}
\label{ssec:manakov}

A particularly important factorizable problem has $\lambda_j = 1$ for
all $j$, so that $\Lambda$ is the identity matrix ${\cal I}$.  In
this case, which we will call \emph{Special Case B}, all of the
interaction strengths $g_{jj'}$ have the value $\sigma g$, and the
atomic polarizabilities $p_j$ are all equal.

The equation of motion~(\ref{eqn:separableEom}) with $\Lambda = {\cal
        I}$ and $V=0$ was first studied by Manakov \cite{manakov1974}
and is now known as the Manakov equation.  We will extend this
terminology by also calling Eq.~(\ref{eqn:separableEom}) with
$\Lambda = {\cal I}$ and nonzero potential $V$ the Manakov equation.

The Manakov Case, i.e., Special Case B, is of considerable physical
interest.  Provided that the atoms are not too close to resonance,
the $p_j$'s are to a good approximation equal~\cite{roberts}.  The
interaction strengths are nearly equal in two-component $^{87}$Rb
condensates~\cite{myatt1997,hall1998}.  As a result, the dynamics of
these condensates are reasonably well described by the Manakov
equation with $s=2$.

Three-component $^{23}$Na condensates with hyperfine spin $F=1$ were
first studied by the MIT group \cite{stamper1998,stenger1998}.  For
$F=1$ spinor condensates, the interaction strengths $g_{jj'}$ are
identical and there are no incoherent couplings if $l_0$ and $l_2$
are equal, where $l_{\cal F}$ is the $s$-wave scattering length for
two colliding atoms with total hyperfine spin ${\cal F}$
\cite{HoTL1998,ohmi1998}.  Since the difference $l_2 - l_0$ is small
compared to $l_0$ for $^{23}$Na \cite{stenger1998,burke1998}, it is a
reasonable approximation use the Manakov equation with $s=3$ to model
the three-component condensates produced by the MIT group, at least
for the initial stage of the time evolution.

For the Manakov case, Eq.~(\ref{eqn:VAl}) reduces to
\begin{equation}
\vert {\bf A}_l \vert^2 = {{\sigma p}\over{4g}} e_l^2\,,
\label{eqn:AlManakov}
\end{equation}
where $l\in\{1,2,\ldots,D\}$.  This shows that $\sigma p$ must be
positive for there to be a solution to Eq.~(\ref{eqn:AlManakov}).
Thus, for the remainder of the paper, when we discuss Special Case B,
we will assume that $\sigma p > 0$.  For convenience, let
\begin{equation}
a_l\equiv{1\over 2}\sqrt{{{\vert p \vert}\over  g}} e_l
\end{equation}
for $l=1,2,\ldots ,D$.  Equation~(\ref{eqn:AlManakov}) is then simply
$\vert {\bf A}_l \vert = a_l$.

The optical potentials $V_j$ are all equal to $V$ for Special Case B.
Equation~(\ref{eqn:potentialPsi2}) shows that
\begin{equation}
V = -\sigma gn + \hbar\Omega,
\label{eqn:potentialdensity}
\end{equation}
where $n\equiv|\bm{\psi}|^2$ is the total condensate number
density.  The total density is independent of time and varies
sinusoidally with position.  The maxima of $n$ are located at the
potential minima for $\sigma=+1$.  In contrast, for $\sigma = -1$,
the maxima of $n$ are located at the maxima of the potential.  This
leads to an obvious instability, as pointed out by Bronski~\textit{et
        al.}~\cite{carr2001d} for the single-component case.
Accordingly, for the remainder of the paper, we will limit our
attention to the case $\sigma = +1$ whenever we study a Case B
problem \cite{footnote2}.  Since we have already assumed that $\sigma
p > 0$, this means that for Case B the atomic polarizability $p$ will
be taken to be positive throughout the remainder of the paper.

The condition~(\ref{eqn:transfCond}) is particularly simple for the
Manakov equation: $P$ can be any unitary matrix. Since $\Lambda$ is
the identity matrix, $T(P)=P$ and $\bm{\zeta}$ is a unitary
transformation of $\bm{\psi}$. Unitary transformations of solutions
to the Manakov equation with an external potential have been studied
elsewhere in one spatial
dimension~\cite{bradleyRM2005,deconinck2004}.  The transformation
given by Eqs.~(\ref{eqn:transformation})--(\ref{eqn:transfCond})
generalizes that work to problems in which the $\lambda_j$'s are not
all identical, as well as to higher spatial dimensions.

\subsection{Factorizable Equations of Motion for Two-Component
Condensates with $\bm{p_1=-p_2}$}
\label{ssec:oppositeSign}

Consider a two-component BEC with factorizable equations of motion.
Recall that $\lambda_1$ and $\lambda_2$ have unit modulus and are
real.  As a result, there are four possibilities: (i)
$\lambda_1=\lambda_2=1$, (ii) $\lambda_1=\lambda_2=-1$, (iii)
$\lambda_1=-\lambda_2=1$ and (iv) $-\lambda_1=\lambda_2=1$.  Case (i)
has already been discussed: it is the Manakov case with $s=2$.  The
equations of motion for case (ii) are unchanged if we reverse the
signs of $\lambda_1$, $\lambda_2$ and $p$, and so case (ii) is
identical to case (i).  In precisely the same way, case (iv) is
equivalent to case (iii).  In this section, we will study case (iii).

The case in which a two-component BEC with factorizable equations of
motion has $\lambda_1=-\lambda_2=1$ will be referred to as
\emph{Special Case C} or the
\emph{factorizable-with-opposite-polarizabilities} (FOP) case.  In
this case, the atomic polarizabilities $p_1$ and $p_2$ have opposite
signs and the interaction matrix
\begin{equation}
M = \sigma g\begin{pmatrix}\:\:\:1 &-1\cr -1 &\:\:\:1\cr\end{pmatrix}.
\label{eqn:rotated}
\end{equation}
If $\sigma$ is positive, atoms in the same condensate component repel
each other and atoms in different condensate components attract.  The
situation is reversed if $\sigma$ is negative.  Finally, note that by
switching the labels of the two components if necessary, we can
arrange for $\sigma p$ to be negative.  We will always take $\sigma
p$ to be negative when we discuss Special Case~C.

For Special Case C, the invertible matrix
\begin{equation}
P(\Delta) \equiv
\begin{pmatrix}
\cosh\Delta &-\sinh\Delta\cr -\sinh\Delta & \cosh\Delta\cr
\end{pmatrix},
\label{eqn:PDelta}
\end{equation}
satisfies the condition~(\ref{eqn:transfCond}) for arbitrary real
$\Delta$.  Equations~(\ref{eqn:transformation}) and
(\ref{eqn:transformation2}) with $P=P(\Delta)$ and real $\Delta$
therefore defines a $P$-set.

\subsection{Two-Component Condensates with a Non-singular Interaction
Matrix}
\label{ssec:nonsingular}

In the special cases discussed so far, the interaction matrix $M$ was
singular.  The cancellation method, however, does yield solutions
even if $\det M$ is nonzero.  To illustrate this point, let us
consider a two-component condensate in one dimension with $\det M\ne
0$.

It follows from Eq.~(\ref{eqn:conditionNE}) that
\begin{equation}
A_{10} A^\ast_{11} = A_{20} A^\ast_{21} = 0\,.
\end{equation}
$A_{11}$ and $A_{21}$ cannot both vanish because the potential
coefficients $V_{j1}$ are nonzero.  If both $A_{11}$ and $A_{21}$ are
nonzero, $A_{10} = A_{20} = 0$, $\Omega_1 = \Omega_2 = 0$, and
\begin{equation}
\psi_j = A_{j1}\cos (k_1 x) e^{-i\omega_1 t}
\end{equation}
for $j=1,2$.  A
solution of this form exists only if the equations
\begin{equation}
V_{j1} =
\sum_{j'=1}^2 g_{jj'} \vert A_{j'1}\vert^2
\label{eqn:Invert}
\end{equation}
have a solution for $A_{11}$ and $A_{21}$.  The
equations~(\ref{eqn:Invert}) have a solution if and only if
\begin{equation}
\chi_1\equiv (g_{22} V_{11} - g_{12} V_{21})/\det M > 0
\label{eqn:chi1}
\end{equation}
and
\begin{equation}
\chi_2\equiv (g_{11} V_{21} - g_{21} V_{11})/\det M >0.
\label{eqn:chi2}
\end{equation}
If the inequalities~(\ref{eqn:chi1}) and (\ref{eqn:chi2}) hold,
$A_{j1}=\sqrt \chi_j$ yields a solution.

We next turn to the case in which only one of the $A_{j1}$'s is
nonzero.  The equations~(\ref{eqn:Invert}) have a solution with
$A_{11}=0$ if
\begin{equation}
{{p_1}\over g_{12}} = {p_2\over g_{22}} > 0.
\label{eqn:condition1}
\end{equation}
On the other hand, Eqs.~(\ref{eqn:Invert}) have a solution with
$A_{21}=0$ if
\begin{equation}
{{p_1}\over g_{11}} = {p_2\over g_{21}} > 0.
\label{eqn:condition2}
\end{equation}
If the condition~(\ref{eqn:condition2}) is satisfied, we can switch
the labeling of the two condensate components, yielding
Eq.~(\ref{eqn:condition1}).  It is therefore sufficient to consider
the case in which the condition~(\ref{eqn:condition1}) holds.  In
this case, $A_{21} = \sqrt{V_{11} / g_{12}}$ and $A_{20}=A_{11}=0$.
It follows that $\psi_1 = A_{10}e^{-i\Omega_1 t}$ and $\psi_2 =
A_{21} e^{-i(\omega_1 + \Omega_2) t}\cos (k_1 x)$.
Equation~(\ref{eqn:OmegaA}) becomes $\hbar\Omega_j= g_{j1}\vert
A_{10}\vert^2$, and, without loss of generality, we may take $A_{10}$
to be real.  We conclude that the two-component order parameter is
given by
\begin{equation}
\psi_1 = a_0e^{-i g_{11}a_0^2 t/\hbar}
\label{eqn:solnD1s2a}
\end{equation}
and
\begin{equation}
\psi_2 = \sqrt{V_{11} \over g_{12}}
e^{-i(\omega_1 + g_{21}a_0^2/\hbar) t}\cos k_1 x \,,
\label{eqn:solnD1s2b}
\end{equation}
where $a_0 \equiv A_{10}$ is an arbitrary real constant.

\section{Application of Analytical Techniques to Select Cases}
\label{sec:cases}

We will now construct solutions to the mean-field equations of
motion~(\ref{eqn:GP}) using the exact analytical methods developed in
the preceding section.  We will start with the simplest cases as an
introduction to the application of our solution methods, and to make
connections with the relatively simple solutions to be found in the
literature.  We will then move on to progressively more rich and
complex problems with higher dimensions and/or more condensate
components than have previously been considered.

\subsection{Solutions on a One Dimensional Optical Lattice}
\label{ssec:1D}

It is convenient to orient the $x$ axis along ${\bf k_1}$, so that
${\bf k}_1 = k_1 \hat x$. Since $\omega_0\ne \omega_1$,
Eq.~(\ref{eqn:conditionNE}) applies with $l=0$ and $l'=1$.

\subsubsection{One-Component Condensates}

We will begin with the simplest case, $D = s = 1$.  For $s = 1$,
Eq.~(\ref{eqn:conditionNE}) reduces to $ g_{11} A_{10}A^\ast_{11} =
0$.  It follows that $A_{10}$ and/or $A_{11}$ must vanish.
Equation~(\ref{eqn:VA}) reduces to $V_{11} = g_{11} \vert
A_{11}\vert^2$. Since $V_{11}$ has been assumed to be nonzero,
$A_{11}$ cannot vanish, and hence $A_{10}=0$.
Equation~(\ref{eqn:OmegaA}) then shows that $\Omega_1=0$.  There is a
solution of the form of~(\ref{eqn:solution}) only if $ g_{11}$ and
$V_{11}$ have the same sign. If this is the case, we have the
solution
\begin{equation}
\psi_1 = \sqrt{ {V_{11}\over
 g_{11}} } \cos (k_1 x) e^{-i\omega_1 t}
\label{eqn:D1n1}
\end{equation}
previously found by Bronski~\textit{et
        al.}~\cite{carr2001c,carr2001d}.  Note that there is a
PC~solution only if the spatially averaged condensate density
$\langle n_1\rangle$ happens to be $V_{11}/(2g_{11})$.  Although this
is a very restrictive condition, this simple case is nevertheless
useful in the development of nonlinear band theory~\cite{carr2005d}.

\subsubsection{Two-Component Condensates}
\label{sssec:1s2d}

We only found a single stationary solution for a one-component
condensate in 1D.  For two-component condensates of the Manakov and
FOP types, we find a much larger parameter space of solutions,
including nonstationary solutions.

We briefly touched on solutions for two-component condensates in one
dimension in Section~\ref{ssec:nonsingular}.  In obtaining the
solution given by Eqs.~(\ref{eqn:solnD1s2a}) and
(\ref{eqn:solnD1s2b}), we did not use our assumption that $\det M \ne
0$.  Moreover, the condition~(\ref{eqn:condition1}) holds for both
case B and case C.  As a result, the solution is \emph{also} valid
for cases B and C.  In both cases, the solution takes the form
$\bm{\psi} = \bm{\psi}_\ast^{(1)}$, where
\begin{equation}
\bm{\psi}_\ast^{(1)} \equiv e^{-i\sigma g a_0^2\Lambda t/\hbar}
\begin{pmatrix}a_0\cr a_1 \cos (k_1 x)
e^{-i\omega_1 t}\cr\end{pmatrix}\,.
\label{eqn:original}
\end{equation}

The solutions for $s = 2$ constructed to this point are stationary,
and have previously been obtained by Deconinck {\it et
        al.}~\cite{deconinck2003}. Let us now consider the Manakov
and FOP cases B and C.  We will demonstrate that for these two cases
there are nonstationary solutions in the same $P$-set as
Eq.~(\ref{eqn:original}), where a $P$-set is the set of solutions
connected by a matrix transformation of the order parameter as
described in Sec.~\ref{sssec:P-set}; in the Manakov case, the
transformation is just a unitary transformation.

For Case B, Eq.~(\ref{eqn:factCondNe}) becomes ${\bf A}_0\cdot {\bf
        A}_1^\ast = 0$. By changing the phase of $\psi_j$ if
necessary, we can arrange for $A_{j0}$ to be real for $j=1,2$. We can
arrange for $A_{11}$ to be real by changing the zero of time if
needed.  It then follows that $A_{21}$ is real as well, and so the
vectors ${\bf A}_0$ and ${\bf A}_1$ have real components. Recalling
that $\vert {\bf A}_1 \vert = a_1$, we obtain ${\bf A}_0 =
a_0(\cos\theta, \sin\theta)^T$ and ${\bf A}_1 = a_1
(-\sin\theta,\cos\theta)^T$, where $a_0$ and $\theta$ are arbitrary
real constants.  The corresponding order parameter is
\begin{equation}
\bm{\psi}({\bf r},t) = e^{-i ga_0^2t/\hbar} [{\bf A}_0 + {\bf A}_1
\cos(k_1x)e^{-i\omega_1 t}]
\label{eqn:solutionD1s2}
\end{equation}
by Eqs.~(\ref{eqn:solution}) and (\ref{eqn:OmegaA0}).  Recasting
this solution, we have $\bm{\psi}({\bf r},t) =
P(\theta)\bm{\psi}_\ast^{(1)} ({\bf r},t)$, where
\begin{equation}
P(\theta)\equiv\begin{pmatrix}\cos\theta &
-\sin\theta\cr \sin\theta
&\:\:\:\:\cos\theta\cr\end{pmatrix}
\label{eqn:Ptheta}
\end{equation}
is a unitary matrix. The solution $\bm{\psi}({\bf r},t)$, which has
been previously described by Bradley~\textit{et
        al.}~\cite{bradleyRM2005}, is therefore a unitary
transformation of the stationary solution~(\ref{eqn:original}).  If
$\sin(2\theta)$ is nonzero, it is a nonstationary solution because
the condensate component densities
\begin{eqnarray}
n_1 &=& \vert\psi_1\vert^2 = a_0^2 \cos^2\theta +
a_1^2 \sin^2\theta\cos^2 (k_1 x)\nonumber\\
&&- a_0 a_1 \sin(2\theta) \cos (k_1 x) \cos(\omega_1 t)
\end{eqnarray}
and
\begin{eqnarray}
n_2 &=& \vert\psi_2\vert^2 = a_0^2 \sin^2\theta + a_1^2
\cos^2\theta\cos^2 (k_1 x)\nonumber\\
&&+ a_0 a_1 \sin (2\theta)\cos (k_1 x) \cos (\omega_1 t)
\end{eqnarray}
oscillate in time with period $T=2\pi/\omega_1$.  We conclude that by
performing unitary transformations of the single solution
$\bm{\psi}_\ast^{(1)}$, we generate an uncountably infinite $P$-set
of stationary and nonstationary solutions.

What is the physical meaning of the
solution~(\ref{eqn:solutionD1s2})?  The external potentials $V_1$ and
$V_2$ coincide and are equal to $V = - {1\over 4} p e_1^2 \cos^2 (k_1
x)$, and so the potential minima occur at the points $x = q \pi/k_1$,
where $q$ is any integer.  We divide the lattice of potential minima
into two sublattices: sublattice 1 with even $q$, and sublattice 2
with odd $q$. The total condensate density $n = n_1 + n_2$ does not
vary in time, and its maxima occur at the points $x = q \pi/k_1$,
where $q$ is any integer.  Suppose for the sake of specificity that
$a_0$ is positive and that $\pi/2 < \theta < \pi$. At time $t=0$, the
maxima of $n_1$ are on sublattice 1, while at time $t=T/2$, the
maxima of $n_1$ are on sublattice 2.  At time $t=T$, the maxima of
$n_1$ are again on sublattice 1.  The maxima of $n_2$ also oscillate
between sublattices 1 and 2, but the oscillations of $n_2$ lag those
of $n_1$ by half a period, ensuring that $n$ is time-independent.

The spatial average of the total number density $\langle n \rangle$
is $a_0^2 + {1\over 2} a_1^2$.  If $\langle n \rangle$ is given,
there is a solution of the form~(\ref{eqn:solutionD1s2}) if and only
if $\langle n \rangle \ge {1\over 2} a_1^2$.  In constrast to the
single-component case, we obtain a solution not just for a single
value of $\langle n \rangle$, but for a whole rangle of $\langle n
\rangle$ values.

This concludes our discussion of the relation between the solutions
obtained using the general formalism of Sec.~\ref{sec:rigorous} and
the existing literature for one dimension. Let us now turn to the
novel FOP case, Special Case C. We can again arrange for the vectors
${\bf A}_0$ and ${\bf A}_1$ to be real.
Equations~(\ref{eqn:factCondNe}) and (\ref{eqn:VAl}) have the
solution ${\bf A}_0 = a_0 (\cosh\Delta, -\sinh\Delta)^T$ and ${\bf
        A}_1 = a_1 (-\sinh\Delta, \cosh\Delta)^T$ valid for arbitrary
real $a_0$ and~$\Delta$.  Since $\hbar\Omega = \sigma g a_0^2$, the
condensate component order parameters are
\begin{eqnarray}
\psi_1 &=& e^{-i\sigma g a_0^2t/\hbar}\left[a_0\cosh\Delta \right.\nonumber\\
&&\left.- a_1\sinh\Delta\cos(k_1 x)e^{-\omega_1 t}\right]
\end{eqnarray}
and
\begin{eqnarray}
\psi_2 &=& e^{i\sigma g a_0^2t/\hbar}\left[-a_0\sinh\Delta\right.\nonumber\\
&&\left. + a_1\sinh\Delta\cos(k_1 x)e^{-\omega_1 t}\right].
\end{eqnarray}
In contrast to the solution~(\ref{eqn:solutionD1s2}) we constructed
for the Manakov case, the component densities $n_1$ and $n_2$
oscillate in phase between the two sublattices, and the total density
$n$ oscillates in time as well.  Since the spatial average of the
total number density $\langle n \rangle = \cosh(2\Delta)(a_0^2 +
{1\over 2} a_1^2)$, we obtain solutions provided that $\langle n
\rangle \ge {1\over 2} a_1^2$.

The vector order parameter may be written
\begin{equation}
\bm{\psi} =
\exp(-i\Omega \Lambda t)P(\Delta)
\exp(i\Omega \Lambda t)\bm{\psi}_\ast^{(1)}\,,
\label{eqn:caseCtransform}
\end{equation}
where $P(\Delta)$ is defined by Eq.~(\ref{eqn:PDelta}).  Once again,
the nonstationary solution can be constructed by transforming the
stationary solution~(\ref{eqn:original}) and we have an uncountably
infinite $P$-set.

\subsubsection{Three-Component Condensates}

We will not carry out the analysis for all possible cases for three
components.  However, we will touch on two new features that arise
when we go from two components to three.

For two-component condensates governed by the Manakov equations of
motion, we showed that the cancellation method only yields solutions
in which the oscillations of the two components between the
sublattices are $180^\circ$ out of phase.  Three-component
condensates have an additional degree of freedom, and this leads to
solutions with a wide range of relative phases.

We will restrict our attention to Case B and to solutions with
$A_{10} = A_{20} = A_{30} > 0$ and nonzero coefficients $A_{j1}$. By
changing the zero of time if needed, we can arrange for $A_{11}$ to
be real and positive. Set $A_{j1} = \vert A_{j1}\vert e^{i\phi_j}$
for $j=1,2,3$ and note that $\phi_1 = 0$.  The
condition~(\ref{eqn:factCondNe}) gives
$A_{11}+A_{21}+A_{31}=0$. Recall that $\vert {\bf A}_1 \vert^2 =
A_{11}^2+A_{21}^2+A_{31}^2 = a_1^2$.  Clearly, there are solutions in
which both $A_{21}$ and $A_{31}$ are real.  In solutions of this
type, two of the components oscillate in phase with one another and
the other component is $180^\circ$ out of phase.

In addition to these solutions, there is a solution for any $\phi_2$
and $\phi_3$ satisfying the conditions
\begin{equation}
0 < \phi_2 < \pi < \phi_3 < 2\pi
\end{equation}
and
\begin{equation}
0 < \phi_3 - \phi_2 < \pi\,.
\end{equation}
The condensate component densities are
\begin{eqnarray}
\vert \psi_j \vert^2 &=& A_{j0} ^2 + \vert A_{j1}
\vert^2 \cos^2 (k_1 x)\nonumber\\
&& + 2 A_{j0} \vert A_{j1} \vert
\cos (k_1 x) \cos(\omega_1 t -\phi_j).
\end{eqnarray}
Thus, a wide variety of phase relationships among the condensate
component densities are possible for $s=3$.  If $s$ is increased
still further, the range of possible phase relationships grows
rapidly.

For three-component condensates, it is possible for the interaction
matrix $M$ to be singular and to have rank greater than one.  Even
though the equations of motion are not factorizable, the cancellation
method developed at the outset of Section~\ref{sec:rigorous} can be
applied to yield solutions for problems of this type.  We will
illustrate this with a one-dimensional example.

Suppose the eigenvalues of $M$ are $\Upsilon_1\ne 0$, $\Upsilon_2\ne
0$ and $\Upsilon_3 = 0$, and let the $\bm{\mu}_i$'s be the associated
real eigenvectors.  We also set
\begin{equation}
{\bf V}_1 \equiv (V_{11}, V_{21},
V_{31})^T
\end{equation}
and
\begin{equation}
{\bf w}_{ll'} \equiv (A_{1l}A_{1l'}^\ast,
A_{2l}A_{2l'}^\ast, A_{3l}A_{3l'}^\ast)^T
\end{equation}
for $l,l' = 0,1$. The
conditions~(\ref{eqn:conditionNE}) and~(\ref{eqn:VA}) are then
\begin{equation}
M{\bf w}_{01}=0\label{eqn:w01}
\end{equation}
and
\begin{equation}
M {\bf w}_{11}={\bf V}_1\,.
\label{eqn:potCond}
\end{equation}
Equation~(\ref{eqn:potCond}) has a solution only if ${\bf V}_1$
is a linear combination of $\bm{\mu}_1$ and $\bm{\mu}_2$. Suppose
this is indeed the case, so that ${\bf V}_1 = \tilde v_{11}
\bm{\mu}_1 + \tilde v_{12} \bm{\mu}_2\,$.  The equation $M\bm{\xi} =
{\bf V}_1$ has the solutions
\begin{equation}
\bm{\xi}
= (\tilde v_{11}/\Upsilon_1) \bm{\mu}_1 + ({\tilde v_{12}/\Upsilon_2})
\bm{\mu}_2 + \tilde\xi_3 \bm{\mu}_3\,,
\end{equation}
where $\tilde\xi_3$ is an arbitrary real constant.  We can set
\begin{equation}{\bf w}_{11} = (\vert A_{11} \vert^2, \vert A_{21}
\vert^2, \vert A_{31} \vert^2)^T =
\bm{\xi}=(\xi_1,\xi_2,\xi_3)^T
\end{equation}
if $\xi_j\ge 0$ for $j=1,2,3$.  Suppose that the $\xi_j$'s are in
fact all positive. Then $A_{j1} = \sqrt{\xi_j} \exp(i\phi_j)$, where
$\phi_j$ is real.  Equation~(\ref{eqn:w01}) shows that
\begin{eqnarray}
{\bf w}_{01} &=& (A_{10}A_{11}^\ast, A_{20}A_{21}^\ast,
A_{30}A_{31}^\ast)^T \nonumber\\
&&= C{\bm\mu}_3=C(\mu_{31}, \mu_{32}, \mu_{33})^T,
\end{eqnarray}
where $C$ is an arbitrary nonzero complex constant, and so $A_{j0} =
C \mu_{3j}\exp(i\phi_j)/\sqrt{\xi_j}$ for $j=1,2,3$.  Since $A_{j0}$
and $A_{j1}$ are both proportional to $\exp(i\phi_j)$, the order
parameter $\psi_j$ is proportional to $\exp(i\phi_j)$ as well, and we
may set $\phi_j=0$ for $j=1,2,3$ without loss of generality.  The
$\Omega_j$'s can be readily obtained from Eq.~(\ref{eqn:OmegaA}).  We
conclude that the order parameter for the $j$th condensate component
is
\begin{equation}
\psi_j = \xi_j^{-1/2} e^{-i\Omega_j t} \left[\xi_j + C
\mu_{3j}\cos(k_1 x)e^{-i\omega_1 t}\right]
\end{equation}
for $j=1,2,3$.  By
changing the zero of time if necessary, we can arrange for $C$ to be
real and positive.  The density of the $j$th condensate component is
then
\begin{eqnarray}
n_j&=&\xi_j + {{C^2\mu_{3j}^2}\over {\xi_j}}\cos^2(k_1 x) \nonumber\\
&&+ 2C\mu_{3j}\cos(k_1 x)\cos(\omega_1 t),
\end{eqnarray}
which shows that the oscillations of each pair of condensate
components are either in phase or $180^\circ$ out of phase. The phase
and amplitude of the oscillations of the components between
sublattices 1 and 2 are determined by the nature of ${\bm\mu}_3$, the
eigenvector of the interaction matrix $M$ with eigenvalue zero.

\subsection{Solutions on a Square Optical Lattice}
\label{ssec:2D}

For $D = 2$, the optical lattice is formed by two standing light
waves with orthogonal wave vectors.  We take the $x$ axis to lie
along $\bf{k}_1$ and the $y$ axis to lie along $\bf{k}_2$.  For $k_1
\ne k_2$, the optical lattice has rectangular symmetry, while for
$k_1=k_2\equiv k$, we obtain a square optical lattice.  In this
section, we will study solutions on the square optical lattice.
Solutions on the rectangular optical lattice will be discussed in
Section~\ref{ssec:2Drect}.

\subsubsection{One-Component Condensates}

The $A_{jl}$'s must satisfy Eq.~(\ref{eqn:conditionNE}) for $l=0$ and
$l'=1,2$.  They must also satisfy Eq.~(\ref{eqn:conditionE}) with
$l=1$ and $l'=2$. For $s=1$, these conditions reduce to $A_{10}
A_{11}^\ast = A_{10} A_{12}^\ast = 0$ and $\Re (A_{11} A_{12}^\ast) =
0$, respectively.  Equation~(\ref{eqn:VA}) has a solution only if
$p_1$ and $ g_{11}$ have the same sign.  We assume that this is the
case.  Since $\vert A_{1l} \vert^2 = V_{1l}/ g_{11} > 0$ for $l=1,2$,
both $A_{10}$ and $\Omega_1$ must vanish.  We can arrange for
$A_{11}$ to be real and positive.  $A_{12}$ is then imaginary.  The
condensate wave function is thus
\begin{equation}
\psi_1 = {1\over 2}
\sqrt{{p_1}\over { g_{11}}}
\left[e_1 \cos (k x) \pm i e_2 \cos (ky) \right] e^{-i\omega t},
\label{eqn:solution2D1s}
\end{equation}
where $\omega \equiv \hbar k^2 / 2\mu$. This is a stationary solution
on the square optical lattice with potential
\begin{equation}
V\equiv V_1 = -{1\over 4}p_1[e_1^2\cos^2 (kx) + e_2^2\cos^2 (ky)]
\end{equation}
and time-independent condensate density $n_1 = -V_1/g_{11}$.  A
PC~solution therefore exists only if the spatially averaged
condensate density $\langle n_1\rangle$ is precisely $p_1(e_1^2 +
e_2^2)/(8g_{11})$.

\begin{figure}[t]
\begin{center}
\includegraphics{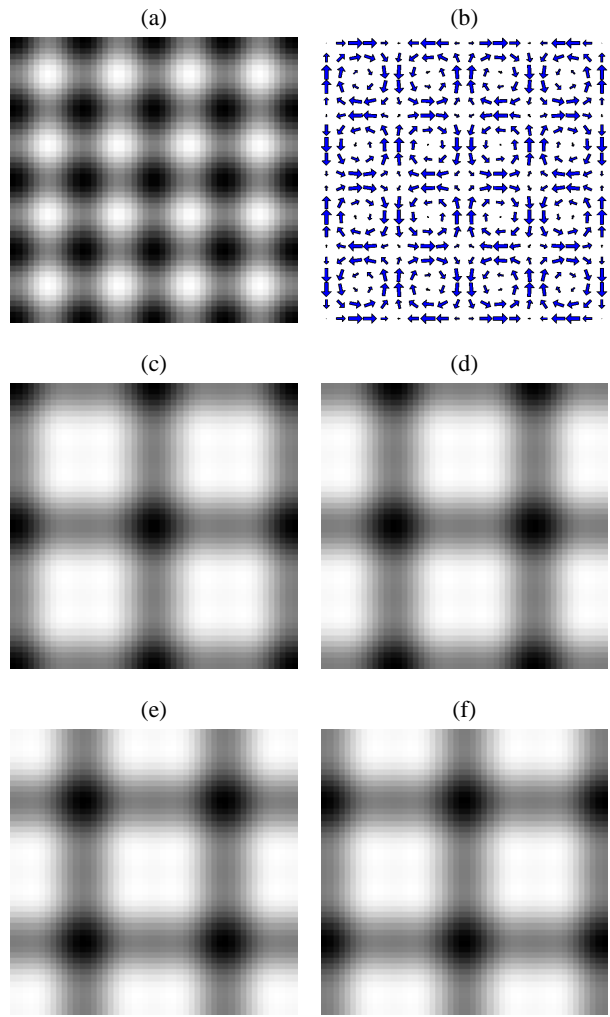}
\caption{\label{fig1} (Color online) \emph{Solutions on a square
                optical lattice}: (a) Gray scale plot of the square
        optical lattice potential $V(x,y)$.  Regions of low (high)
        potential are shown in black (white).  (b) The current
        density for the one-component
        solution~(\ref{eqn:solution2D1s}) with the upper sign.  The
        direction (size) of the arrows indicates the direction
        (magnitude) of the current flow.  This solution is a
        vortex-anti-vortex array.  (c)--(f) Gray scale plots of the
        density of the first component $n_1(x,y)$ for the
        two-component solution to the Manakov case with $a_0=a_1$,
        $\rho=1$ and $\theta = 3\pi/4$.  The plots are for times
        $t=T/8$ [Panel~(c)], $t=3T/8$ [Panel~(d)], $t=5T/8$
        [Panel~(e)] and $t=7T/8$ [Panel~(f)].  Regions of high (low)
        $n_1$ are shown in black (white).  Each plot shows the region
        with $-\lambda\le x \le\lambda$ and $-\lambda\le y
        \le\lambda$.}
\end{center}
\end{figure}

Although the solution~(\ref{eqn:solution2D1s}) has previously been
constructed by Hai~\textit{et al.}~\cite{hai2004}, its physical
interpretation has not yet been discussed.  The solution is a
vortex-antivortex lattice [see Fig.~\ref{fig1}, Parts (a) and (b)],
and so the condensate is flowing even though its density is not
time-dependent.  Each square of side $\lambda/2$ with a potential
maximum at its center and potential minima at its corners is occupied
by a vortex or antivortex.  The cores of the vortices and
antivortices are located at the potential maxima, where the
condensate density is zero.  As shown in Fig.~\ref{fig1}(b), the
vortices and antivortices are arranged in a checkerboard pattern.

The mean-field equations of motion~(\ref{eqn:GP}) are time-reversal
invariant: if $\bm{\psi}({\bf r}, t)$ is a solution, then so is the
time-reversed state $\bm{\psi}^\ast({\bf r}, -t)$.  Using our method
of solution, we found both the solution with the upper sign in
Eq.~(\ref{eqn:solution2D1s}) and its time-reversed version, the
solution with the lower sign.

\subsubsection{Two-Component Condensates}

We have now finished establishing contacts between the literature and
the solutions obtained using the formalism of
Section~\ref{sec:rigorous}.  To the best of our knowledge, the
solutions found from this point on are new.

The analysis for two-component condensates on a square optical
lattice runs parallel to that given for two components on a
one-dimensional optical lattice, and so only the final results will
be given. If $\det M\ne 0$, there is a solution of the form
\begin{equation}
\psi_j = (\vert A_{j1}\vert \cos kx \pm i \vert A_{j2}\vert \cos ky)
e^{-i\omega t}
\end{equation}
for $j=1,2$, provided that Eq.~(\ref{eqn:VA}) has
a solution for the $ A_{jl}$'s.  Each of the two condensate
components moves in a vortex-antivortex lattice in this solution.  If
the condition~(\ref{eqn:condition1}) holds, on the other hand, we
have a solution with
\begin{equation}
\psi_1 = a_0 e^{-i g_{11}a_0^2 t/\hbar}
\end{equation}
and
\begin{eqnarray}
\psi_2 &=&
{1\over 2} \sqrt{{p_1}\over {g_{12}}}
(e_1 \cos k_1 x \pm i e_2 \cos ky)\nonumber\\
&&e^{-i(\omega + g_{21}a_0^2/\hbar) t} ,
\end{eqnarray}
where $a_0 = A_{10}$ is an arbitrary real constant.  For both Case B
and Case C, this is a valid solution and $\bm{\psi} =
\bm{\psi}_\ast^{(2)}$, where
\begin{eqnarray}
\bm{\psi}_\ast^{(2)} &\equiv& e^{-i\sigma g a_0^2\Lambda t/\hbar}\nonumber\\
&&\times\begin{pmatrix}a_0\cr
 ( a_1 \cos kx \pm i a_2 \cos ky) e^{-i\omega t}\cr\end{pmatrix}.
\label{eqn:original2Ds2}
\end{eqnarray}

\begin{figure}[t]
\begin{center}
\includegraphics{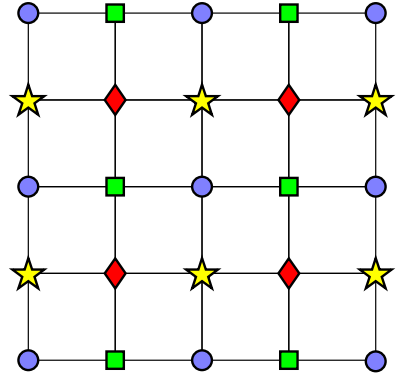}
\caption{\label{fig2} (Color online) Division of the lattice of
        potential minima into four square sublattices.  The lattice
        of potential minima are the vertices of the grid.  The
        vertices in sublattices 1, 2, 3 and 4 are indicated by
        circles, squares, diamonds and stars, respectively.  The
        origin is at the center of the figure.}
\end{center}
\end{figure}

There are nonstationary solutions if the problem is factorizable.
For the Manakov Case B, we have an uncountably infinite $P$-set: the
solution $\bm{\psi} = P(\theta)\bm{\psi}_\ast^{(2)}$ is valid for
arbitrary real $\theta$.  The densities of the two components of the
condensate are time-dependent in this solution if $\sin(2\theta)$ is
nonzero. For example,
\begin{eqnarray}
n_1(x,y,t) &=& a_0^2\cos^2\theta
+ a_1^2 \sin^2\theta[\cos^2 (k x) + \rho^2 \cos^2 (k y)] \nonumber\\
&&- a_0 a_1 \sin (2\theta)[\cos (k x) \cos (\omega t) \nonumber\\
&&+\rho \cos (k y) \sin (\omega t)],
\label{eqn:2D2sdensity}
\end{eqnarray}
where $\rho\equiv\pm e_2/e_1$.  The solution with $\rho = - e_1/e_2$
is simply the time-reversed version of the solution with $\rho =
e_1/e_2$, and so we may restrict our attention to the case $\rho >
0$.  The external potentials $V_1$ and $V_2$ coincide and are equal
to $V = -{1\over 4}p(e_1^2\cos^2 kx + e_2^2\cos^2 ky)$.  The
potential minima occur at the points $(x, y) = {\lambda\over 2} (q_1,
q_2)$, where $q_1$ and $q_2$ are integers and $\lambda$ is the
optical wavelength.  We divide the lattice of potential minima into
four square sublattices with lattice spacing $\lambda$, as shown in
Fig.~\ref{fig2}.

Although $n_1$ and $n_2$ are time-dependent, the total condensate
density $n$ does not vary in time, and its maxima occur at the
potential minima.  The time evolution of the density of the first
component, as described by Eq.~(\ref{eqn:2D2sdensity}), is
illustrated in Fig.~\ref{fig1}(c)--(f).  Let $T=2\pi/\omega$ be the
period and suppose for the sake of specificity that $a_0$ is positive
and that $\pi/2 < \theta < \pi$.  At time $t=T/8$, the maxima of
$n_1$ are on sublattice 1 [Fig.~\ref{fig1} (c)]. One quarter period
later, the maxima of $n_1$ are on sublattice 2 [Fig.~\ref{fig1} (d)].
They are on sublattice 3 at time $t=5T/8$ [Fig.~\ref{fig1} (e)] and
sublattice 4 at time $t=7T/8$ [Fig.~\ref{fig1} (f)].  Finally, the
maxima of $n_1$ return to sublattice 1 at time $t=9T/8$.  The maxima
of $n_2$ also oscillate among the sublattices, but the oscillations
of $n_2$ lag those of $n_1$ by half a period.

For the FOP Case C, we obtain a $P$-set of solutions by transforming
$\bm{\psi}_\ast^{(2)}$: explicitly, $\bm{\psi} = e^{-i\Omega\Lambda
        t}P(\Delta)e^{i\Omega\Lambda t}\bm{\psi}_\ast^{(2)}$ is a
solution for arbitrary real $\Delta$.  The external potentials are
$V_1 = - V_2 = V$.  Suppose for the sake of specificity that $p$ is
positive.  The minima of $V_1$ and the maxima of $V_2$ then occur at
the lattice of points $(x, y) = {\lambda\over 2} (q_1, q_2)$, where
$q_1$ and $q_2$ are integers and $\lambda$ is the optical wavelength.
We divide this lattice into the same four square sublattices as we
did for Case B. As time passes, the maxima of $n_1$ move periodically
among the four sublattices, just as they do for Case B.  In Case C,
however, the oscillations of component 2 are in phase with those of
component 1, and the total condensate density $n$ varies in time as a
result.  This is analogous to what we found for Case~C in one
dimension.

For the solutions just discussed, the spatial average of the total
number density $\langle n \rangle = a_0^2 + {1\over 2} (a_1^2 +
a_2^2)$ for the Manakov case, while $\langle n\rangle
=\cosh(2\Delta)[a_0^2 + {1\over 2} (a_1^2 + a_2^2)]$ for the FOP
case.  In both cases, we obtain solutions provided that $\langle n
\rangle\ge {1\over 2} (a_1^2 + a_2^2)$.

\subsection{Solutions on Rectangular Optical Lattices}
\label{ssec:2Drect}

We now turn to the case in which $D=2$ and $k_1\ne k_2$, i.e., to the
rectangular optical lattice.  No solutions of the form of
Eq.~(\ref{eqn:solution}) exist for a single-component condensate on a
rectangular optical lattice.  For a two-component condensate,
PC~solutions are obtained only for the Manakov Case B, and so we will
confine our attention to that case.  Choosing $l=0$ and $l'=1,2$ in
Eq.~(\ref{eqn:factCondNe}), we observe that if ${\bf A}_0$ is
nonzero, ${\bf A}_1$ and ${\bf A}_2$ must be parallel.
Equation~(\ref{eqn:factCondNe}) with $l=1$ and $l'=2$ then shows that
either ${\bf A}_1$ or ${\bf A}_2$ must vanish and, hence,
Eq.~(\ref{eqn:VAl}) cannot be satisfied for both $l=1$ and $l=2$. It
follows that ${\bf A}_0 = 0$.  We can take the components of ${\bf
        A}_1$ to be real without loss of generality.  Since $\vert
{\bf A}_1\vert = a_1$, we may set ${\bf
        A}_1=a_1(\cos\theta,\sin\theta)^T$.  We can arrange for the
components of ${\bf A}_2$ to be real through a change in the zero of
time.  Because ${\bf A}_1\cdot{\bf A}_2=0$ and $\vert {\bf A}_2\vert
= a_2$, it follows that ${\bf A}_2=\pm
a_2(-\sin\theta,\cos\theta)^T$.  Equation~(\ref{eqn:OmegaA0}) shows
that $\Omega=0$, and hence we have the solution given by
\begin{eqnarray}
\psi_1&=& a_1\cos\theta\cos(k_1 x) e^{-i\omega_1 t} \nonumber\\
&&\mp a_2\sin\theta\cos(k_2 y) e^{-i\omega_2 t}
\label{eqn:rec1}
\end{eqnarray}
and
\begin{eqnarray}
\psi_2&=& a_1\sin\theta\cos(k_1 x) e^{-i\omega_1 t} \nonumber\\
&&\pm a_2\cos\theta\cos(k_2 y) e^{-i\omega_2 t},
\label{eqn:rec2}
\end{eqnarray}
where the angle $\theta$ is arbitrary.  Equations~(\ref{eqn:rec1})
and (\ref{eqn:rec2}) define a $P$-set of solutions since
$\bm{\psi}=P(\theta)(a_1\cos(k_1 x) e^{-i\omega_1 t}, \pm a_2\cos(k_2
y) e^{-i\omega_2 t})^T$.

\begin{figure}[t]
\begin{center}
\includegraphics{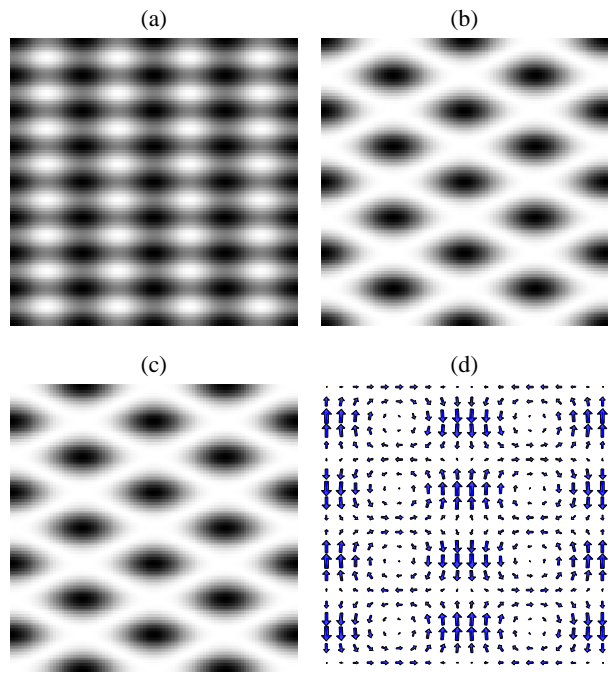}
\caption{\label{fig3} (Color online) \emph{Two-component condensate
                on a rectangular optical lattice with
                $\lambda_1=2\lambda_2$}: (a) Gray scale plot of the
        2D optical potential $V(x,y)$.  Regions of low (high)
        potential are shown in black (white).  (b)--(c) Density of
        the first component $n_1(x,y,t)$ at times $t=0$ and $T/2$,
        respectively; regions of high (low) $n_1$ are shown in black
        (white).  (d) Current density of the first condensate
        component at time $t=T/4$, illustrating the flow from
        sublattice $A$ to sublattice $B$.  The direction (size) of
        the arrows indicate the direction (magnitude) of the current
        flow.  Each of the four plots shows the region with
        $-\lambda_1\le x \le\lambda_1$ and $-\lambda_2\le y
        \le\lambda_2$.}
\end{center}
\end{figure}

Equations~(\ref{eqn:rec1}) and (\ref{eqn:rec2}) give a nonstationary
solution with temporal period $T=2\pi/|\omega_2-\omega_1|$ for $0 <
\theta < \pi/2$.  The time evolution of this solution can be
understood as follows.  The optical potential $V=-{1\over
        4}p[e_1^2\cos^2 (k_1 x) + e_2^2\cos^2 (k_2 y)]$ has minima at
the points $(x, y) = \pi (q_1/k_1, q_2/k_2)$, where $q_1$ and $q_2$
are integers.  Divide the lattice of potential minima into two
sublattices: sublattice $A$ with even $q_1+q_2$ and sublattice $B$
with odd $q_1+q_2$.  For the solution given by Eqs.~(\ref{eqn:rec1})
and~(\ref{eqn:rec2}) with the lower signs and $0 < \theta < \pi/2$,
the maxima of $n_1$ are initially on sublattice $A$, as illustrated
in Fig.~\ref{fig3}(b) for $a_0=a_1$ and $\theta = \pi/4$.  Half a
period later, maxima of $n_1$ are on sublattice $B$ [see
Fig.~\ref{fig3}(c)].  The maxima of $n_1$ are on sublattice A once
again at time $t=T$. The second component oscillates between the two
sublattices in the same way, but its oscillations lag those of the
first component by half a period.

In the limit that $k_1$ and $k_2$ coincide, the period of oscillation
$T$ tends to infinity and we obtain a stationary solution on the
square optical lattice.  In this solution, the maxima of $n_1$ reside
on one sublattice and the maxima of $n_2$ are on the other.

\subsection{Solutions on the Simple Cubic Optical Lattice}
\label{ssec:3D}

For condensates with two or more components on three dimensional
optical lattices, the set of solutions of the
form~(\ref{eqn:solution}) is prohibitively large.  Therefore, we will
make a number of simplifying assumptions and will limit ourselves to
giving examples of solutions.  The stationary solutions we will
discuss all have a high degree of symmetry.

For $D=3$, the optical lattice is formed by three standing waves with
orthogonal wave vectors.  The $x_l$ axis will be taken to lie along
${\bf k}_l$ for $l=1,2,3$.  We will confine our attention to the case
in which the three standing waves have the same wavelength $\lambda$,
so that the lattice of potential minima is a simple cubic (SC)
lattice with lattice spacing $\lambda/2$.  We will further simplify
the problem by restricting our attention to the Manakov Case B and by
assuming that the $e_l$'s coincide.  To simplify the notation, set
$k\equiv k_1=k_2=k_3$, $\omega\equiv\hbar k^2/2m$, $e\equiv
e_1=e_2=e_3$, $f_l({\bf r}) \equiv \cos({\bf k}_l\cdot{\bf r})$ for
$l=1,2,3$, and $a\equiv\sqrt{p/g} e/2$.

From Section~\ref{sec:rigorous}, we know that
\begin{eqnarray}
{\bm\psi} &=& \exp( -ig|{\bf A}_0|^2 t/\hbar)\nonumber\\
&&\times\left[{\bf A}_0 +
({\bf A}_1 f_1 + {\bf A}_2 f_2 + {\bf A}_3 f_3)e^{-i\omega t} \right]
\label{eqn:3Dsolution}
\end{eqnarray}
is a solution to the mean-field equations of motion if
\begin{equation}
|{\bf A}_l|= a \quad {\rm and} \quad {\bf A}_0^\ast\cdot{\bf A}_l=0
\label{eqn:conditions1}
\end{equation}
for $l=1,2,3$ and
\begin{equation}
\Re({\bf A}_1^\ast\cdot{\bf A}_2) =
\Re({\bf A}_2^\ast\cdot{\bf A}_3) =
\Re({\bf A}_3^\ast\cdot{\bf A}_1) = 0.
\label{eqn:conditions2}
\end{equation}
There is no solution to Eqs.~(\ref{eqn:conditions1})
and~(\ref{eqn:conditions2}) for $s=1$, and so we will only consider
condensates with two or more components.

\begin{figure}[t]
\begin{center}
\includegraphics{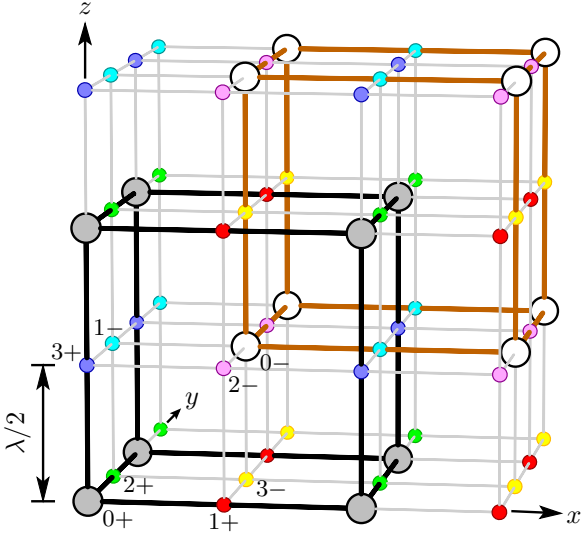}
\caption{\label{fig4} (Color online) \emph{Sublattices of the simple
                cubic lattice}: The vertices of the grid are sites of
        the lattice of potential minima.  The gray sites belong to
        the SC sublattice $0+$, while the white sites belong to the
        SC sublattice $0-$.  The gray and white sites together make
        up the BCC sublattice $0$.  Sites of all eight SC sublattices
        are labelled at the corners of the cube in which $x$, $y$ and
        $z$ all lie between 0 and $\lambda/2$.  The sites of SC
        sublattices $1+$, $1-$, $2+$, $2-$, $3+$ and $3-$ are colored
        red, cyan, green, magenta, blue and yellow, respectively.}
\end{center}
\end{figure}

For brevity, the lattice of potential minima will be referred to as
\lq\lq the lattice."  The lattice can be divided into eight simple
cubic sublattices with lattice spacing $\lambda$. The vector
\begin{equation}
{\bf f} \equiv (f_1, f_2, f_3)
\end{equation}
takes on a different value on each of these sublattices.  The lattice
can also be divided into four body-centered cubic (BCC) sublattices.
Each of these BCC sublattices is the union of two simple cubic
sublattices with ${\bf f}$'s that sum to zero.  In
Table~\ref{labels}, we assign labels to each of the eight simple
cubic sublattices and to each of the four BCC sublattices.  These
sublattices are illustrated in Fig.~\ref{fig4} and will play an
important role in our examples.
\begin{table}[tbh]
\caption{\label{labels}Labeling of the sublattices of the simple
cubic lattice}
\begin{ruledtabular}
\begin{tabular}{lcc}
{\bf f}&SC sublattice label&BCC sublattice label\\
\hline
(1,1,1)   &$0+$&0\\
(-1,-1,-1)&$0-$&0\\
(-1,1,1)  &$1+$&1\\
(1,-1,-1) &$1-$&1\\
(1,-1,1)  &$2+$&2\\
(-1,1,-1) &$2-$&2\\
(1,1,-1)  &$3+$&3\\
(-1,-1,1) &$3-$&3\\
\end{tabular}
\end{ruledtabular}
\end{table}

All of the stationary solutions we will discuss have at least one of
the two symmetries we will now define.  If, after a certain lattice
translation, the density $n_j({\bf r})$ is unchanged by a rotation of
$90^\circ$ about the $x$, $y$ and $z$ axes for $j=1,2,\ldots ,s$,
then we say that a solution has \emph{four-fold rotational symmetry}.
Note that the lattice translation could depend on the condensate
component index $j$ and could be the null translation.  If a solution
has four-fold rotational symmetry, the $x$, $y$ and $z$ directions
are equivalent, and so this symmetry is a type of discretized
isotropy.  On the other hand, if for each pair $(j, j')$ there is a
sequence of lattice translations or a series of rotations of
$90^\circ$ about the $x$, $y$ or $z$ axes that maps $n_j({\bf r})$
onto $n_{j'}({\bf r})$, then we say that the solution has
\emph{component symmetry}.  Intuitively speaking, the $s$ components
of the condensate all play the same role in a solution with component
symmetry.

\subsubsection{Two-Component Condensates}

For $s=2$, all solutions of the form~(\ref{eqn:3Dsolution}) are
stationary because Eqs.~(\ref{eqn:conditions1})
and~(\ref{eqn:conditions2}) do not have a solution with nonzero ${\bf
        A}_0$.  Both examples of solutions we give will therefore be
stationary solutions.

One possible solution with four-fold rotational symmetry is given by
\begin{equation}
\psi_1
= {1\over\sqrt{3}} a (f_1+f_2+f_3) e^{-i\omega t} \equiv \psi_1^{(rot)}
\end{equation}
and
\begin{equation}
\psi_2 = \sqrt{2\over 3} a (f_1+ e^{2\pi
        i/3} f_2+ e^{4\pi i/3} f_3) e^{-i\omega t}\equiv
\psi_2^{(rot)}\,.
\end{equation}
The maxima of the total density are located at
the potential minima.  A straightforward analysis reveals that $n_1$
has its maxima on the BCC sublattice~0 and that the maxima of $n_2$
reside on BCC sublattices 1, 2 and 3.  This solution does not possess
component symmetry.

Component 1 is at rest since the phase of $\psi_1$ is independent of
position.  In contrast, the flow of component 2 is fascinating: it
flows in a three-dimensional vortex lattice of great beauty, as we
will now demonstrate.

Consider the cube ${\cal C}$ in which $x$, $y$ and $z$ range between
0 and $\lambda/2$.  Each of the eight corners of the cube belong to a
different simple cubic sublattice (see Fig.~\ref{fig4}).  The second
component of the condensate flows along six of the twelve edges
of the cube.  Specifically, component 2 flows from the site $1+$ to
the site $3-$, and then to sites $2+$, $1-$, $3+$ and $2-$ before
returning to site $1+$.  There is no mass current along the remaining
six edges of the cube.

The cyclic flow of component 2 suggests that there is a vortex line
within the cube, and this is fact the case: a vortex line has its
core along the cube diagonal that joins the $0+$ site to the $0-$
site.  To establish this, we will begin by considering the behavior
of $\psi_2$ close to the line $x=y=z$.  Let
\begin{equation}
\hat{\bf e}_1'=\sqrt{2\over 3}\left(\hat x -{1\over 2} \hat y -{1\over
        2} \hat z\right),
\end{equation}
\begin{equation}
\hat{\bf e}_2'=\sqrt{1\over 2}(\hat
y - \hat z),
\end{equation}
and
\begin{equation}
\hat{\bf e}_3'=\sqrt{1\over 3}(\hat x + \hat
y + \hat z).
\end{equation}
The vectors $\hat{\bf e}_1'$, $\hat{\bf e}_2'$ and
$\hat{\bf e}_3'$ form an orthonormal triad with $\hat{\bf e}_3'=
\hat{\bf e}_1' \times\hat{\bf e}_2'$.  We introduce the new
coordinates $x_i'= \hat{\bf e}_i'\cdot{\bf r}$, where $i=1,2,3$.  On
the line $x=y=z$, both $x_1'$ and $x_2'$ vanish and $x_3'$ is
arbitrary.  Rewriting $\psi_2$ in terms of the new coordinates and
expanding for small $x_1'$ and $x_2'$, we obtain
\begin{equation}
\psi_2\cong
-ka\sin\left({{kx_3'}\over{\sqrt{3}}}\right)(x_1'+ix_2')e^{-i\omega t}.
\label{eqn:vortex}
\end{equation}
Equation~(\ref{eqn:vortex}) shows that there is a vortex line
with its core along the line $x=y=z$.  The direction of the vector
$\hat{\bf e}_3'$ and the direction of the current flow around the
vortex core are related by the right hand rule; for brevity, we will
say that the vortex line is \emph{oriented} along the vector
$\hat{\bf e}_3'$.

We have just shown that there is a vortex line within the cube ${\cal
        C}$ with its core along the cube diagonal, and that the
vortex line is oriented along the vector $\hat{\bf e}_3'$.  To
determine the nature of the flow throughout space, first note that
$\psi_2$ is invariant under the transformation $x\to -x$, i.e., it is
invariant under reflection about the $y-z$ plane.  Naturally,
$\psi_2$ is also invariant under the reflections $y\to -y$ and $z\to
-z$.  These invariances give the flow within the cubical region in
which $x$, $y$ and $z$ range from $-\lambda/2$ to $+\lambda/2$.
Since $\psi_2$ is invariant under the lattice translation ${\bf r}\to
{\bf r} + (q_1\hat x + q_2\hat y+q_3\hat z)\lambda$ for all integers
$q_1$, $q_2$ and $q_3$, the nature of the flow in the whole of space
can now be inferred.

The following picture emerges from this analysis.  There is a vortex
core along every line in the BCC sublattice~0 that joins an infinite
chain of nearest-neighbor sites.  At these sites, the density of the
second component is at a maximum, its current density is zero and the
potential is at a minimum.  Each vortex core also passes thorough a
chain of neighboring potential maxima which alternate with the
potential minima.  At the potential maxima, the density of the second
component of the condensate $n_2$ is zero.  In this way, the energy
of the vortex array is minimized.  Every vortex line is oriented
along one of the following four vectors: $\hat x+ \hat y + \hat z$,
$\hat x- \hat y - \hat z$, $-\hat x + \hat y - \hat z$ or $-\hat x-
\hat y + \hat z$.

The solution with
\begin{equation}
\psi_1 = {a\over \sqrt{2}} (f_1+ e^{i\pi /4}
f_2+ e^{-i\pi /4} f_3) e^{-i\omega t}
\end{equation}
and
\begin{equation}
\psi_2 = {a\over
        \sqrt{2}} (f_1- e^{-i\pi /4} f_2 - e^{i\pi /4} f_3)
e^{-i\omega t}
\end{equation}
has component symmetry but not four-fold
rotational symmetry: The $x$ direction is not equivalent to the $y$
and $z$ directions, and $n_1$ and $n_2$ differ only by a translation
through the distance $\lambda/2$ along the $x$ axis.  The maxima of
$n_1$ are on BCC sublattice~0, while the maxima of $n_2$ are on BCC
sublattice 1.

A natural question to ask is whether, for a given $s$, there is a
solution with both four-fold rotational \emph{and} component
symmetries.  The answer to this question is \lq\lq no" for both $s=2$
and 3, as we show in Appendix~\ref{app:3D}.

\subsubsection{Three-Component Condensates}

For $s=3$, the stationary solution given by $\psi_1 =
\psi_1^{(rot)}$, $\psi_2 = \sqrt{\phi}\, \psi_2^{(rot)}$ and $\psi_3
= \sqrt{1-\phi}\, \psi_2^{(rot)}$ with $0 < \phi < 1$ has four-fold
rotational symmetry but does not have component symmetry.  Next,
consider the stationary solution with
\begin{equation}
\psi_j = {2\over 3} a
\left(f_1+f_2+f_3 - {3\over 2} f_j\right) e^{-i\omega t},
\end{equation}
for $j=1,2,3$.  The maxima of $n_j$ are on the $j$th sublattice and
each component of the condensate is at rest.  This solution does not
have four-fold rotational symmetry since the $x_j$ direction is
special for condensate component $j$.  However, it does have
component symmetry.  To see this, consider an arbitrary pair of
indices $(l,l')$ with $l\ne l'$. Let $l''$ be the integer belonging
to the set $\{1,2,3\}$ that differs from both $l$ and $l'$.  A
$90^\circ$ rotation about the $x_{l''}$ axis interchanges $f_l$ and
$f_{l'}$, and so maps $n_l$ onto $n_{l'}$.

A nonstationary solution that illustrates just how complex the
solutions for $s=3$ can be is given by
\begin{equation}
\psi_1={a\over \sqrt{2}}(f_1+f_2+if_3)e^{-i(\omega+\Omega) t},
\end{equation}
\begin{equation}
\psi_2={a\over 2}\left[\sqrt{2}+(-f_1+f_2+if_3)e^{-i\omega
        t}\right]e^{-i\Omega t},
\end{equation}
and
\begin{equation}
\psi_3={a\over
        2}\left[\sqrt{2}-(-f_1+f_2+if_3)e^{-i\omega
        t}\right]e^{-i\Omega t},
\end{equation}
where $\Omega=ga^2/\hbar$.  The density of the first condensate
component is time-independent and its maxima are on BCC sublattices~0
and~3.  The density maxima of component 2 are on simple cubic
sublattice~1+ at time $t={T\over{2\pi}}\tan^{-1}(1/2)\equiv\tau$, on
simple cubic sublattice~2+ at $t=T/2-\tau$, on simple cubic
sublattice~$1-$ at $t=T/2+\tau$, and are on simple cubic
sublattice~$2-$ at time $t=T-\tau$.  At time $T+\tau$, the maxima of
$n_2$ have returned to simple cubic sublattice~1+.  The motion of
condensate component~3 is identical to that of the second component,
except that the oscillations of $n_3$ lag those of $n_2$ by half a
period.

\subsubsection{Four-Component Condensates}

The solution space for four-component condensates is very large.  To
see this, consider an arbitrary set of orthonormal vectors with real
components in four dimensions, $\{\hat {\bm\epsilon}_0, \hat
{\bm\epsilon}_1, \hat {\bm\epsilon}_2, \hat {\bm\epsilon}_3\}$.
Setting ${\bf A}_0=a_0 \hat {\bm\epsilon}_0$ and ${\bf A}_l=a \hat
{\bm\epsilon}_l$ for $l=1,2,3$, we obtain a solution to
Eqs.~(\ref{eqn:conditions1}) and (\ref{eqn:conditions2}) for
arbitrary nonnegative real numbers $a_0$.  One such solution is given
by
\begin{equation}
\psi_0 = {1\over 2}\left[-a_0 + a(f_1+f_2+f_3) e^{-i\omega
        t}\right ]e^{-ig a_0^2 t/\hbar}
\label{eqn:soln3D4sA}
\end{equation}
and
\begin{eqnarray}
\psi_j &=& {1\over 2}
\left[a_0 + a(f_1+f_2+f_3-2f_{j-1}) e^{-i\omega t}\right]\nonumber\\
&&\times e^{-ig a_0^2 t/\hbar}
\label{eqn:soln3D4sB}
\end{eqnarray}
for $j=2$, 3, and 4.

For two- and three-component condensates, no stationary solution of
the form~(\ref{eqn:3Dsolution}) has both four-fold rotational and
component symmetries.  Such a solution does exist for four-component
condensates, however.  The solution is given by
Eqs.~(\ref{eqn:soln3D4sA}) and~(\ref{eqn:soln3D4sB}) with $a_0=0$.
In this solution, the maxima of $n_j$ are on sublattice $j-1$ and
each of the four condensate components is at rest.  To see that the
solution has four-fold rotational symmetry, note that $n_1$ is
unchanged by a rotation of $90^\circ$ about the $x$, $y$ and $z$
axes.  After a translation through ${1\over 2}\lambda \hat x_{j-1}$,
the density $n_j$ becomes $n_1$ and so is unchanged by a rotation of
$90^\circ$ about the $x$, $y$ and $z$ axes for $j=2$, 3, 4.

As we have seen, $n_j$ can be mapped onto $n_1$ by a primitive
lattice translation for $j=2$, 3 and 4.  On the other hand, $n_1$ is
mapped onto $n_{j'}$ by a translation through ${1\over 2}\lambda \hat
x_{j'-1}$ for $j'=2$, 3, and 4. It follows that, for each pair $(j,
j')$, there is a sequence of at most two primitive lattice
translations that carries $n_j$ onto $n_{j'}$, and so the solution
has component symmetry.

For $a_0>0$, Eqs.~(\ref{eqn:soln3D4sA}) and~(\ref{eqn:soln3D4sB})
describe a nonstationary solution.  In this solution, the density
maxima of the first component of the condensate are on the simple
cubic sublattice~$0-$ at time $t=0$ and are on the simple cubic
sublattice~$0+$ at $t=T/2$.  For $j=2$, 3 and 4, the maxima of $n_j$
are on the simple cubic sublattice~$j+$ at time $t=0$ and are on the
simple cubic sublattice~$j-$ at $t=T/2$.  The entire condensate
returns to its initial state at time $t=T$.

\section{Nonlinear Stability}
\label{sec:stability}

\subsection{Dimensionless mean-field equations}
\label{ssec:stab_dimensionless_mfe}
Our numerical investigations of the stability of selected
solutions to the mean-field equations (\ref{eqn:GP}) are performed
with a dimensionless form of the equations, using dimensionless
position, time, and potential-energy variables,
\begin{equation}
\xivec
=
\rvec / x_0
\,,\quad
\tau = t / t_0
\,,\quad\text{and}\quad
\tilde V_j = V_j / E_0
\,,
\end{equation}
defined in terms of units $x_0$, $t_0$, and $E_0$.
For optical potentials of the form (\ref{eqn:potential2}), we
also define dimensionless potential-strength coefficients
\begin{equation}
\tilde V_{jl}
=
V_{jl}/E_0
\,.
\end{equation}
The length and time units, $x_0$ and $t_0$, are related to
the energy unit $E_0$ by
\begin{equation}
x_0 = \hbar/\sqrt{mE_0}
\quad\text{and}\quad
t_0 = \hbar/E_0
\,.
\label{eq:x0_t0}
\end{equation}

The elements $g_{jj'}$ of the interaction matrix
must also be put into dimensionless
form, but first we note that they may be renormalized, depending on the
dimensionality of the optical lattice.
If there is narrow harmonic transverse confinement for
one-dimensional and two-dimensional optical lattices, the appropriate
forms in all dimensions are~\cite{carr2005c}
\begin{equation}
\begin{aligned}
g_{jj'}^{(3)}
&=
\frac{4\pi\hbar^2}{m}
a_{jj'}\,,
&\quad
g_{jj'}^{(2)}
&=
\left(
\frac{8\pi\hbar^3\omega_z}{m}
\right)^{\!1/2}
a_{jj'}\,,
\\
\multispan4\hfill$
\text{and}\quad
\displaystyle
g_{jj'}^{(1)}
=
2\hbar\omega_{\perp}
a_{jj'}
\,,
$\hfill
\label{eq:gjjpd}
\end{aligned}
\end{equation}
where $a_{jj'}$ is the low-energy $s$-wave scattering length for
species $j$ and $j'$, the superscript on $g_{jj'}$ is $D$, the
dimensionality of the optical lattice, and the confining potentials
are characterized by the angular frequencies $\omega_z$ in two
dimensions and $\omega_{\perp}$ in one dimension.

We may choose the elements $\tilde g_{jj'}$ of the dimensionless
form of the interaction matrix
to be typically of order one, so that the
elements of the
scattering-length matrix and the dimensional interaction matrix
decompose as
\begin{equation}
a_{jj'}
=
a
\tilde g_{jj'}
\quad\text{and}\quad
g_{jj'}^{(D)}
=
g^{(D)}
\tilde g_{jj'}
\,,
\end{equation}
where $a$ and $g^{(D)}$ are scalar, dimensional factors.  The latter
is the same as the $g$ appearing in \eqref{eqn:factor}, but with
the possible need for renormalization in lower dimensions explicitly
indicated by the superscript.  The dimensionless form of $g^{(D)}$ is
\begin{equation}
\tilde g^{(D)}
=
\frac{
g^{(D)}
}{
E_0 x_0^D
}
\,.
\label{eq:gd_def}
\end{equation}
Its values, which follow from
Eqs.~(\ref{eq:x0_t0})--(\ref{eq:gd_def}), are
\begin{equation}
\begin{aligned}
\tilde g^{(3)}
&=
\frac{4\pi}{\hbar}
\sqrt{mE_0}
a\,,
&\qquad
\tilde g^{(2)}
&=
\left(
\frac{8\pi m\omega_z}{\hbar}
\right)^{\!1/2}
a\,,
\\
\multispan4\hfill$
\text{and}\quad
\displaystyle
\tilde g^{(1)}
=
2
\sqrt{
\frac{
m
}{
E_0
}
}
\omega_{\perp}
a
\,.
$\hfill
\label{eq:gd}
\end{aligned}
\end{equation}

We absorb the square root of this dimensionless scale factor into the
order parameter, which then takes the dimensionless form
\begin{equation}
\tilde\psi_j
=
\sqrt{\tilde g^{(D)}}
x_0^{D/2}
\psi_j
\,.
\label{eq:psij_nodim}
\end{equation}
Thus, for a PC solution, the coefficients appearing in
$\tilde\psi_j$ are related to those in \eqref{eqn:solution} by
\begin{equation}
\tilde A_{jl}
=
\sqrt{\tilde g^{(D)}}
x_0^{D/2}
A_{jl}
\,.
\end{equation}
The normalization of the dimensionless order parameter is given by
\begin{equation}
\sum_{j=1}^s
\int
\bigabs{
\tilde\psi_j(\xivec,\tau)
}^2
\,d\xivec
=
N\tilde g^{(D)}
\quad
\forall
~
\tau
\,,
\label{eq:psij_nodim_norm}
\end{equation}
wherein we see that $\tilde g^{(D)}$ plays a role equivalent to the
number of particles, $N$.  For a PC solution, the
dimensionless mean number density is then
\begin{equation}
\begin{split}
\avg{\tilde n}
&\equiv
x_0^D
\avg n
\\&=
\frac{1}{\tilde g^{(D)}}
\sum_{j=1}^s
\biggl(
\abs{\tilde A_{j0}}^2
+
\frac12
\sum_{l=1}^D
\abs{\tilde A_{jl}}^2
\biggr)
\,,
\end{split}
\label{eq:dimensionless_den}
\end{equation}
where $\avg n$ is given in \eqref{eqn:A_0}.

Finally, the mean-field equations (\ref{eqn:GP}) take
the dimensionless form
\begin{equation}
i
\pd{\tilde\psi_j}{\tau}
=
\biggl[
-\frac12
\nabla_{\xi}^2
+
\biggl(
\sum_{j'=1}^s
\tilde g_{jj'}
\abs{\tilde\psi_{j'}}^2
\biggr)
+
\tilde V_j
\biggr]
\tilde\psi_j
\,.
\label{eq:gpe_nodim}
\end{equation}

\subsection{Details of the calculations}
\label{ssec:stab_details}
Our numerical stability tests use \eqref{eq:gpe_nodim}, propagating a
specified initial condition $\smash{\tilde\psi_j(\xivec,0)}$ forward
in time via a fifth-order Runge-Kutta algorithm with adaptive
step-size control~\cite{footnote4}\nocite{press1993},
the spatial derivatives being calculated in wave-vector space
via a pseudo-spectral method.

We perturb the solution by adding some white noise to the initial
condition before beginning the time propagation.  This is
accomplished for each component of the order parameter by adding to
the real and imaginary parts of each of its Fourier components a
random number from a uniform distribution in the range
$\pm0.5\times10^{-4}$ times the modulus of the largest Fourier
component.

We work within a spatial cell comprising four periods of the optical
lattice (two optical wavelengths) in each of the $D$ dimensions,
applying periodic boundary conditions to that cell.  The spatial
grids contain $n_{\text{grid}}=128$ points for one-dimensional cases
and $n_{\text{grid}}=32$ points in each dimension for two-dimensional
cases.  Thereby we are able to test the stability of solutions against
perturbations having wavelengths ranging from
$2\lambda/n_{\text{grid}}$ to $2\lambda$, where $\lambda$ is the
optical wavelength.

As a measure of the instability of a component of a solution at a
particular instant of time, we use the variance of its Fourier power
spectrum relative to that at $\tau=0$~\cite{carr2005d,carr2005e},
\begin{equation}
\sigma_j(\tau)
=
\sqrt{
\frac{\displaystyle
\sum_{\kappavec}
\bigl[\tilde f_j(\kappavec,\tau)-\tilde f_j(\kappavec,0)\bigr]^2
}{\displaystyle
2\sum_{\kappavec}
\bigl[\tilde f_j(\kappavec,0)\bigr]^2
}}
\,.
\end{equation}
Here $\tilde
f_j(\kappavec,\tau)\equiv\bigabs{\tilde\phi_j(\kappavec,\tau)}^2$,
$\tilde\phi_j(\kappavec,\tau)$ is the Fourier transform of
$\tilde\psi_j(\xivec,\tau)$, and $\kappa_i\equiv 2\pi/\xi_i$.

A solution is deemed to have reached the onset of instability when
each of the $\sigma_j$ has exceeded $0.1$ at least once.  We find
that this criterion correlates nicely with the visual onset of
instability in the graph of the density and works well for a range of
solution types and potential strengths.

The time of onset of instability can be sensitive to many details,
including the amount of added noise, the resolution of the spatial
grid on which the solution is represented, and even details of the
generation of the random deviates and the algorithm used to perform
the fast Fourier transforms, particularly when the solution is stable
for long times.  As well, we expect the lifetime of an experimentally
produced condensate to vary with the level of noise
present.  Consequently, one should not infer from our graphs of
instability-onset time \vs\ solution parameters that the times
represent literal lifetimes that would be observed in any particular
experiment.

However, as will become clear below, there is a fairly well-defined
boundary between unstable solutions and stable solutions, beyond
which the instability-onset times increase extremely rapidly.  The
locations of those boundaries are largely insensitive to details of
the calculations.  We therefore expect the parameter boundaries
delimiting numerically stable solutions to be experimentally
meaningful, in the sense that within the stable regions observed
lifetimes should be at least of order one second.

\subsection{Results of the calculations}
\label{ssec:stab_results}
To make our results more concrete, we have chosen typical
values for the laser wavelengths, $\lambda=800~\text{nm}$ in
all directions,
the trap frequencies, $\omega_z=2\pi\times100~\text{Hz}$ and
$\omega_{\perp}=2\pi\times200~\text{Hz}$, the atomic mass,
$m=87~\text{u}$, and the $s$-wave scattering length,
$a=55~\text{\AA}$.  We will refer to these below as the
``system parameters.''

The results are displayed primarily in recoil units, setting
\begin{align}
E_0
&=
E_R
=
\frac{\hbar^2k_L^2}{2m}
=
\frac{2\pi^2\hbar^2}{m\lambda^2}
\approx
0.172~\mu\text{K}
\times k_B
%2.37\times10^{-30}~\text{J}
\\
\intertext{and}
t_0
&=
t_R
=
\hbar/E_R
\approx
4.44\times10^{-5}~\text{s}
\,,
\label{eq:recoil_units}
\end{align}
where $k_L$ is the laser wave number, and $k_B$ is Boltzmann's constant.
The numerical values are obtained from our chosen system parameters
above.

We present below selected numerical stability analyses for one to three
condensate components in one and two dimensions.  Three-dimensional
cases are not included, for they are too computationally demanding at
this time.

It is straightforward to transform the system-specific values shown
in the figures below, the potential strengths, the instability-onset
times, and the numbers of particles per well, to values appropriate
for alternative choices of the system parameters.  We will elaborate
on this point in \secref{ssec:stab_altsys}, following the
presentation of the results.

\subsubsection{One component on a square lattice}
\label{sssec:stab_1s2d}
We choose the dimensionless interaction parameter and the
dimensionless potential coefficients $\tilde V_{jl}$ to be
\begin{align}
\tilde g_{11} = 1.0
\,,
&&
\tilde V_{11} = \mathcal V/E_R
\,,
&&\text{and}&&
\tilde V_{12} = \mathcal V/E_R
\,,
\end{align}
where the potential-strength parameter $\mathcal V$ can be varied.
Then the dimensionless solution corresponding to
\eqref{eqn:solution2D1s} with the upper sign has coefficients
\begin{align}
\tilde A_{10} = 0
\,,
&&
\tilde A_{11} = \sqrt{\tilde V_{11}}
\,,
&&\text{and}&&
\tilde A_{12} = i\sqrt{\tilde V_{12}}
\,.
\end{align}
Because the spatially constant term is required to vanish, the
particle density, \eqref{eqn:A_0}, is uniquely determined by the
potential-strength parameter.

\begin{figure}[t]
\includegraphics{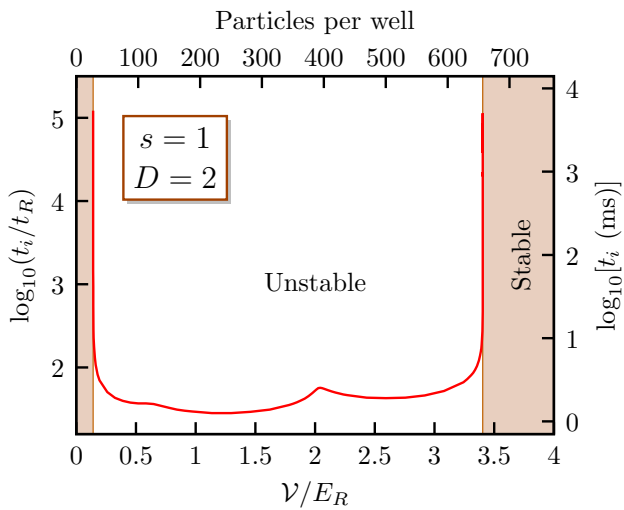}
\caption{\label{fig:instab_1s2d}(Color online)  Instability-onset
	times $t_i$ for a one-component PC solution on
	a square optical lattice.  The dimensionless
        potential-strength parameter $\mathcal V/E_R$ and the
        corresponding particle density are shown on the horizontal
        axes.  The solution is stable at both low and high potential
        strengths.}
\end{figure}

The instability-onset times for this solution are shown as a function
of $\mathcal V$ in \figref{fig:instab_1s2d}.  Two time scales are
included, showing both recoil times and milliseconds, with a maximum
propagation time of several seconds.  The top scale shows the
particle density in particles per well corresponding to the
potential-strength parameter shown on the bottom scale.

While the solution becomes unstable in just a short time over much of
the range of potential strengths shown, it is stable for sufficiently
weak potentials.  Much more surprisingly, it is also stable for
sufficiently strong potentials.  To test whether the solution becomes
unstable again for potentials stronger than that at the boundary near
$3.5E_R$, we performed propagations to
$t\approx1542t_R\approx68~\text{ms}$ of solutions having $\mathcal
V/E_R$ as high as $48$, finding no recurrence of instability.  This
extends well into the Mott insulating regime, beyond the point of
physical relevance of the mean-field equations, as discussed in
\secref{sec:introduction}.

\subsubsection{Two components on a one-dimensional lattice}
\label{sssec:stab_2s1d}
Here we study the stability for the Manakov case, in which the
dimensionless interaction matrix has rank one and has all elements
equal to one.  The potential coefficients we choose are
\begin{equation}
\tilde V_{11} = \tilde V_{21} = \mathcal V/E_R
\,,
\end{equation}
and the coefficients of the dimensionless solution corresponding to
\eqref{eqn:original} are
\begin{equation}
\begin{aligned}
\tilde A_{10} &= \alpha\,,
&&&
\tilde A_{20} &= 0\,,
\\
\tilde A_{11} &= 0\,,&&\text{and}
&
\tilde A_{21} &= \sqrt{\tilde V_{11}}
\,,
\label{eq:2s1d_soln}
\end{aligned}
\end{equation}
where $\alpha$ is a free parameter.
The components of this solution are then mixed using $P(\theta)$ of
\eqref{eqn:Ptheta} with $\theta=\pi/4$ to produce a nonstationary
solution.

\begin{figure}[t]
\includegraphics{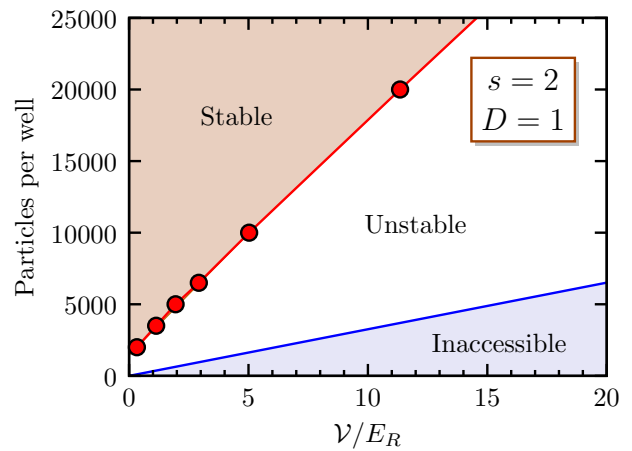}
\caption{\label{fig:instab_2s1d_nv}(Color online) Regions of
	stability and instability for a PC
	solution having two components on a one-dimensional optical
	lattice.  The points denote calculated boundaries between stable
	and unstable behavior, the lines connecting them serving as
        guides to the eye.  The inaccessible region has no
        PC solutions of the form
        (\ref{eq:2s1d_soln}).}
\end{figure}

Since the spatially constant term in the resulting solution is not
required to vanish, we can vary the parameter $\alpha$ to control the
particle density independently of the potential-strength parameter,
giving a two-dimensional domain in which to investigate the stability
of the solution.  As is clear from \figref{fig:instab_1s2d}, the
boundaries of the stable regions are approximated well by the
positions at which the instability-onset times have exceeded about
three hundred times $t_R$.  Consequently, we have used that as
a threshold to define those boundaries for the present case in
scans over the potential-strength parameter at several fixed
values of the particle density.  The resolution of the
potential grid was $0.16E_R$, easily adequate for the graphical
delimitation of the stable regions.  The results are shown as a map
of stable and unstable regions in \figref{fig:instab_2s1d_nv}.

The boundary of the region marked ``inaccessible'' corresponds to the
vanishing of the coefficient $\tilde A_{10}$ of the constant term in
the solution.  Within that region it is impossible to construct a PC
solution of the chosen form, since \eqref{eqn:lengthA0} cannot be
satisfied.  The region marked ``stable'' is that portion of the
parameter space where the criterion for stability is satisfied, and
the region marked ``unstable'' corresponds to parameters for which
the instability-onset time falls below the threshold.  As in
the two-dimensional case shown in \figref{fig:instab_1s2d}, the
solution becomes stable in the weak-potential limit.  However, in
striking contrast to the two-dimensional case, there is no evidence
of a second region of stability at high potential strengths.

In order to verify that apparent absence of stability for deep
potentials, we performed calculations along the boundary of the
inaccessible region with $\mathcal V/E_R$ as high as $48$,
well beyond the highest shown in \figref{fig:instab_2s1d_nv}, finding
only monotonically decreasing instability-onset times~\cite{footnote5}.
This confirms the observation in \figref{fig:instab_2s1d_nv} that
there is no additional stable region at high potential strength.

\subsubsection{Two components on a square lattice}
\label{sssec:stab_2s2d}
As we did in one dimension, here we study the Manakov case, with all
elements of the dimensionless interaction matrix equal to one.  Now
there are four potential coefficients, which we choose to be equal:
\begin{equation}
\tilde V_{11} = \tilde V_{12} = \tilde V_{21} = \tilde V_{22}
=
\mathcal V/E_R
\,.
\end{equation}
The coefficients of the solution corresponding to
\eqref{eqn:original2Ds2} are
\begin{equation}
\begin{aligned}
\tilde A_{10} &= \alpha\,,
&
\tilde A_{20} &= 0\,,
\\
\tilde A_{11} &= \tilde A_{12} = 0\,,
&
\tilde A_{21} &= \sqrt{\tilde V_{11}}\,,&&\text{and}
\\
\tilde A_{22} &= i\sqrt{\tilde V_{11}}
\,,
\label{eq:2s2d_soln}
\end{aligned}
\end{equation}
where $\alpha$ allows us to set the particle density.  Once again
we mix the components using $P(\theta)$ with $\theta=\pi/4$ to obtain
a nonstationary solution.

\begin{figure}[t]
\includegraphics{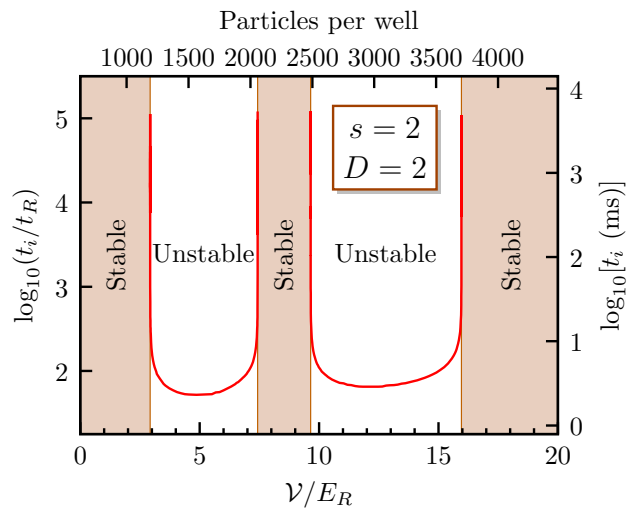}
\caption{\label{fig:instab_2s2d}(Color online) Instability-onset times
	for a two-component PC solution on a
        square optical lattice.  The dimensionless
        potential-strength parameter $\mathcal V/E_R$ and
        corresponding particle density are shown on the horizontal
        axes.  The solution is stable in three regions, having low,
        high, and intermediate potential strengths.}
\end{figure}

Instability-onset times for this solution with $\alpha$ fixed at one
and varying potential strengths are shown in
\figref{fig:instab_2s2d}, where the conventions are similar to those
used for the single-component case in \figref{fig:instab_1s2d}.  As
in that case, regions of stability occur at both low and high
potential strengths, but now a rather striking additional region of
stability appears at intermediate potential strengths, just below
$10$ recoil energies.

\begin{figure}[b]
\includegraphics{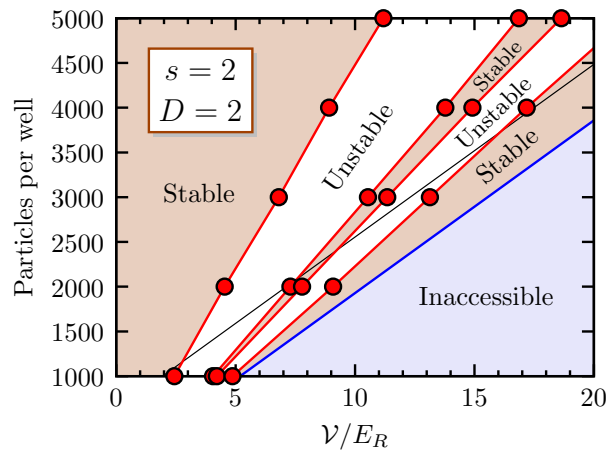}
\caption{\label{fig:instab_2s2d_nv}(Color online) Regions of
	stability and instability for the PC solution
        \protect\eqref{eq:2s2d_soln}, having two components on a
        square optical lattice.  The points denote
        calculated boundaries between stable and unstable behavior,
        the lines connecting them serving as guides to the eye.  The
        inaccessible region has no PC solutions of
        the form \protect\eqref{eq:2s2d_soln}.  The fine line
        parallel to the boundary of the inaccessible region is the
        track followed by the graph of instability-onset times in
        \protect\figref{fig:instab_2s2d}.}
\end{figure}

To explore this behavior in more detail, we map out regions of
stability in the plane of particle density and potential strength in
\figref{fig:instab_2s2d_nv} using the same strategy applied in the
one-dimensional case in \figref{fig:instab_2s1d_nv}.  Three areas of
stability are clearly evident, though the central one is somewhat
narrower than the others.  The fine black line running parallel to
the boundary of the inaccessible region is the track in the
parameter-space plane followed by the graph shown in
\figref{fig:instab_2s2d}.  We extended the search for renewed
instability along this line to $\mathcal V/E_R=48$, limiting the
propagation time to $t\approx1542t_R\approx 68~\text{ms}$, finding no
evidence of further instability beyond the crossover into the stable
region near $\mathcal V/E_R=16$.

\subsubsection{Three components on a one-dimensional lattice}
\label{sssec:stab_3s1d}
We have also tested the stability of a class of PC
solutions having three components.  The dimensionless interaction matrix
then has nine elements, all ones in the Manakov case, and rank equal to one.
The potential coefficients are all chosen to be the same:
\begin{equation}
\tilde V_{11} = \tilde V_{21} = \tilde V_{31}
=
\mathcal V/E_R
\,.
\end{equation}
The coefficients of the components of the dimensionless solution are
all of equal magnitude, those of the spatially constant term being
real:
\begin{equation}
\tilde A_{10}
=
\tilde A_{20}
=
\tilde A_{30}
= \alpha
\,,
\end{equation}
with those of the space-dependent part chosen to have the phases of
the cube roots of unity:
\begin{equation}
\begin{aligned}
\tilde A_{11}
&=
\sqrt{\frac{\tilde V_{11}}{3}}\,,
&\qquad
\tilde A_{21}
&=
\sqrt{\frac{\tilde V_{11}}{3}}
e^{i2\pi/3}\,,\quad\text{and}
\\
\multispan4\hfill$
\displaystyle
\tilde A_{31}
=
\sqrt{\frac{\tilde V_{11}}{3}}
e^{i4\pi/3}
\,.
$\hfill
\label{eq:3s1d_soln}
\end{aligned}
\end{equation}

\begin{figure}[t]
\includegraphics{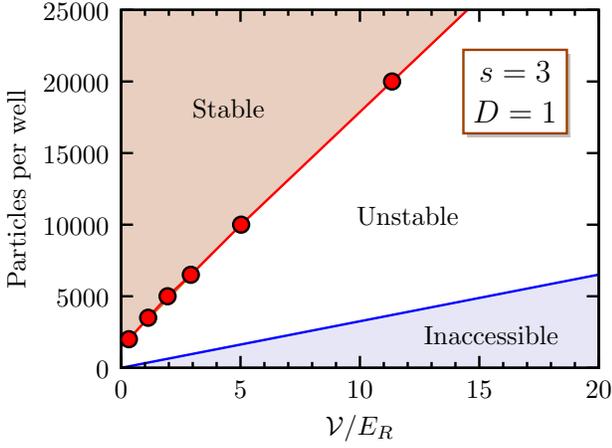}
\caption{\label{fig:instab_3s1d_nv}(Color online) Regions of
	stability and stability for a PC
	solution having three components on a one-dimensional optical
	lattice.  The points denote calculated boundaries between stable
	and unstable behavior, the lines connecting them serving as
        guides to the eye.  The inaccessible region has no
        PC solutions of the form
        \protect\eqref{eq:3s1d_soln}.}
\end{figure}

The map of stable regions in the space of the free parameters of this
solution is shown in \figref{fig:instab_3s1d_nv}, which is visually
indistinguishable from its two-component analog,
\figref{fig:instab_2s1d_nv}, having a region of stability where the
potential is sufficiently weak.  In fact, the coordinates of all the
points plotted on the graph are identical, within the
resolution of the scans.

\subsection{Alternative choices of system parameters}
\label{ssec:stab_altsys}
Those aspects of the results presented above that are dependent on
the system parameters chosen in \secref{ssec:stab_results} are
readily transformed to values corresponding to alternative choices of
those parameters.  Obviously, the potential-strength parameter
$\mathcal V$ is trivially obtained by multiplying the dimensionless
value $\mathcal V/E_R$ by the recoil energy corresponding to any
desired set of system parameters, and the instability-onset time $t_i$ is
easily converted by multiplying it by the ratio $t_R'/t_R$ of the time
units $t_R'$ corresponding to the alternative parameters and
$t_R$ corresponding to our chosen system parameters.

The average number of particles per well is just the mean density
times the well volume,
\begin{equation}
N_{\text{well}}
=
\left(\frac{\lambda}{2}\right)^{\!D}
\avg n
\,.
\end{equation}
From the dimensionless density given in \eqref{eq:dimensionless_den},
we see that this can be expressed in terms of dimensionless
parameters as
\begin{equation}
\begin{split}
N_{\text{well}}
&=
\left(\frac{\lambda}{2x_R}\right)^{\!D}
\frac{1}{\tilde g^{(D)}}
\sum_{j=1}^s
\biggl(
\abs{\tilde A_{j0}}^2
+
\frac12
\sum_{l=1}^D
\abs{\tilde A_{jl}}^2
\biggr)
\,,
\end{split}
\label{eq:nwell_dimensionless}
\end{equation}
with the length unit $x_0$ set to the recoil length
$x_R=\hbar/\sqrt{mE_R}$.

For all of our stability figures, the dimensionless potential-strength
parameter $\mathcal V/E_R$ is a free parameter, and it determines the
values of the dimensionless coefficients $\tilde A_{jl}$ having $l>0$.
For Figures~\ref{fig:instab_2s1d_nv}, \ref{fig:instab_2s2d_nv}, and
\ref{fig:instab_3s1d_nv}, there is one additional free parameter,
$\alpha$, and it determines the values of one or more of the
coefficients $\tilde A_{j0}$.  Thus, for any
given abscissa in Figure~\ref{fig:instab_1s2d} or
\ref{fig:instab_2s2d}, or abscissa and ordinate in
Figure~\ref{fig:instab_2s1d_nv}, \ref{fig:instab_2s2d_nv}, or
\ref{fig:instab_3s1d_nv}, the sum in
\eqref{eq:nwell_dimensionless} is fixed, and the value of
$N_{\text{well}}$ corresponding to an alternative choice of system
parameters can be obtained from that shown on the graph by simply
rescaling the prefactors:
\begin{equation}
N_{\text{well}}'
=
N_{\text{well}}
\left(\frac{2x_R}{\lambda}\right)^{\!D}
\tilde g^{(D)}
\left(\frac{\lambda'}{2x_R'}\right)^{\!D}
\frac{1}{\tilde {g^{(D)}}'}
\,,
\end{equation}
where the unprimed quantities correspond to our choice of system
parameters, and the primed quantities to some alternative choice.

\section{Conclusions}
\label{sec:conclusion}

In this paper, we made a comprehensive study of potential-canceling
(PC) solutions for an $s$ component Bose-Einstein condensate in in a
$D$-dimensional optical lattice.  Studies of specific cases with
small $s$ and $D$, especially in one spatial dimension, have appeared
in the literature.  Our work brings these previous studies together,
generalizes to arbitrary $s$ and $D$, and provides intriguing new
solution types and novel physical interpretations.

Currently, there is a great deal of interest in the a
Berezinskii-Kosterlitz-Thouless phase in Bose-Einstein condensates at
intermediate temperatures in 2D~\cite{krugerP2007}.  In such a phase,
vortex-anti-vortex pairs become bound together, in contrast to the
free vortex proliferation which occurs at high temperatures.
However, this phase is restricted to a truly 2D system, which is
difficult to achieve experimentally.  We have shown that an optical
lattice stabilizes vortex-anti-vortex pairs in the
quasi-two-dimensional case, and that the lattice causes the array to
be tightly packed.

Not only have we presented multicomponent generalizations of 2D
vortex-anti-vortex arrays, but we have also generalized them to three
dimensions.  In 3D, we constructed a solution in which one condensate
component forms a lovely and complex three-dimensional array of
intersecting vortex lines.  As a part of our study of PC solutions in
3D, we gave a thorough treatment of the most highly symmetric
solutions for condensates with two, three and four components.  Our
formalism can also be used to gain insight into complex systems now
experimentally available, such as five-component condensates in 3D
optical lattices.

We studied the stability of PC solutions numerically in 1D and 2D for
one-, two and three-component condensates.  We found three main
results: (1) potential-canceling solutions tend to become stable as
the potential strength is reduced; (2) there is a remarkable
difference between the one-dimensional and two-dimensional solutions,
in that the latter are also stable for deep potentials; and (3) for
two-components in a square optical lattice, there is a fascinating
third region of stability for intermediate-strength potentials.  We
found no evidence of stabilization at high potential strength in one
dimension.

Finally, we mention that the possibility of experimentally realizing
vortex-anti-vortex arrays in a 2D lattice is provided for in the
recent experiments of Sebby-Strabley {\it et
        al.}~\cite{sebbystrabley2006}.  In those experiments, two
polarizations of the lasers used to create the optical lattice
potential are manipulated to create lattices that can be dynamically
controlled on a site-by-site basis.  In this way, one can imagine
creating an array of small ``propellers'' to stir up
vortex-anti-vortex pairs.  The parameter ranges in which such a
procedure would lead to stable structures were determined in our
numerical studies.  To manipulate multicomponent condensates, one can
imagine more advanced versions of such an experiment, in which the
fact that different hyperfine components ``feel'' different lattice
strengths for a given optical wavelength $\lambda$ can be used to
one's advantage.

We thank B. Deconinck, J. N. Kutz, and J. N. Roberts for useful
discussions.  LDC's work was supported by the National Science
Foundation under Grant PHY-0547845 as part of the NSF CAREER program.

\appendix
\section{Proof that the $\bm{\lambda_j}$'s can be rescaled to have
unit modulus}
\label{app:lambda}

For Special Case A, the equations of motion are factorizable and are
given by Eq.~(\ref{eqn:separableEom}).  The elements of the
interaction matrix are $g_{jj'} = \sigma g\lambda_j\lambda_{j'}$ and
the optical potentials are $V_j = \lambda_j V$.  Consider an
associated \lq\lq normalized" problem that is also factorizable.  In
this normalized problem, the elements of the interaction matrix are
$\tilde{g}_{jj'} = \sigma g\tilde{\lambda_j}\tilde{\lambda}_{j'}$ and
the optical potentials are $\tilde{V}_j = \tilde{\lambda}_j V$, where
$\tilde{\lambda}_j\equiv\lambda_j/\vert\lambda_j\vert$ has unit
modulus.  Suppose we have a solution
\begin{equation}
\tilde{\psi}_j =
e^{-i\tilde{\Omega}_j t} \sum_{l=0}^s \tilde{A}_{jl}\cos({\bf
        k}_l\cdot{\bf r}) e^{-i\omega_l t}
\label{eqn:solutionTilde}
\end{equation}
to the normalized problem.  We can then construct a corresponding
solution to the original, unnormalized problem as follows.  We let
$A_{jl}=\tilde{A}_{jl}/\sqrt{\vert\lambda_j\vert}$ and $\Omega_j =
\vert\lambda_j\vert \tilde{\Omega}_j$, and define $\bm{\psi}$ through
Eq.~(\ref{eqn:solution}).  $\bm{\psi}$ is then a solution to
Eq.~(\ref{eqn:separableEom}).  Moreover, $\vert\psi_j\vert^2 =
\vert\tilde{\psi}_j\vert^2/\vert\lambda_j\vert$.  We see that for
each solution $\bm{\tilde{\psi}}$ of the normalized problem, there is
a corresponding solution $\bm{\psi}$ to the original, unnormalized
problem, and that the density of the $j$th condensate component
simply differs by the constant factor $\vert\lambda_j\vert^{-1}$ in
the two problems.  As a result, we may assume without loss of
generality that the $\lambda_j$'s all have unit modulus.

The length of $\tilde{\bf A}_0$ is a free parameter at this point.
However, if the average total density $\langle n \rangle$ is
specified in the original, unnormalized problem, then
Eq.~(\ref{eqn:A_0}) gives
\begin{equation}
\langle n \rangle =
\sum_{j=1}^s
{1\over {\lambda_j}}\left(|\tilde{A}_{j0}|^2
+ {1\over 2}\sum_{l=1}^D |\tilde{A}_{jl}|^2
\right).
\label{eqn:tildeConstraint}
\end{equation}
If Eq.~(\ref{eqn:tildeConstraint}) has a solution, it fixes the value
of $|\tilde{\bf A}_0|^2$.  We conclude that $|\tilde{\bf A}_0|^2$ is
determined if $\langle n \rangle$ is given.

\section{Proof that there are no solutions with both four-fold
rotational and component symmetries for two- and
three-component condensates in 3D}
\label{app:3D}

It was stated in Sec.~\ref{ssec:3D} that there are no solutions with
both four-fold rotational and component symmetries in 3D if the
condensate has two or three components.  Our proof is as follows. Let
$s$ be 2 or 3. The condensate order parameters are
\begin{equation}
\psi_j = \left ( \sum_{l=1}^3 A_{jl}f_l\right)
e^{-i\omega t},
\end{equation}
where $j$ ranges from 1 to $s$ and $f_l\equiv\cos({\bf k}_l\cdot {\bf r})$.
The densities are
\begin{equation} n_j = \sum_{l=1}^3 |A_{jl}|^2 f_l^2  + 2\sum_{1\le
l<l' \le 3}\Re(A_{jl}A_{jl'}^*)f_l f_{l'}. \end{equation} If $n_j$
is to possess four-fold rotational symmetry, the coefficients of $f_1^2$,
$f_2^2$ and $f_3^2$ must be the same. Thus, $|A_{jl}|$ must be
independent of $l$.  If the solution is to have the component
symmetry, on the other hand, $|A_{jl}|$ cannot depend on $j$. It
follows that $|A_{jl}|^2 = a^2/s$ for all $j$ and $l$.

By choosing a phase, we can arrange for $A_{j1}$ to be real for
$j\in\{1,2,\ldots, s\}$.  Let $A_{j2} = a e^{i\alpha_j}/\sqrt{s}$ and
$A_{j3} = a e^{i\beta_j}/\sqrt{s}$, where $ \alpha_j$ and $\beta_j$
are real. Then
\begin{eqnarray}
n_j &=& {{a^2}\over s}\left\{ f_1^2 + f_2^2 + f_3^2 + 2\left[\cos
(\alpha_j) f_1 f_2 \right.\right.\nonumber\\
&&\left.\left. +\cos( \alpha_j-\beta_j) f_2 f_3 + \cos (\beta_j) f_3
f_1 \right]\right\}.
\label{eqn:densities}
\end{eqnarray}
For $n_j$ to have four-fold rotational symmetry, we must have
\begin{equation}
|\cos \alpha_j|=|\cos( \alpha_j-\beta_j)| = |\cos
\beta_j|\,,
\end{equation}
while the condition
\begin{equation}
|\cos \alpha_1| = \ldots = |\cos \alpha_s|\,
\label{eqn:alphas}
\end{equation}
must be satisfied if the solution is to have component symmetry.

Equation~(\ref{eqn:alphas}) must be reconciled with the condition
$\Re({\bf A}_1^*\cdot {\bf A}_2)=0$, i.e.,
\begin{equation}
\sum_{j=1}^s \cos
\alpha_j = 0. \label{eqn:sumAlphas}
\end{equation}
This is not possible for
$s=3$, and so there is no solution with both four-fold rotational and
component symmetries in that case.

The case $s=2$ requires further analysis.
Equation~(\ref{eqn:sumAlphas}) gives $ \alpha_2 = \pi + \sigma_1
\alpha_1$, where $\sigma_1 = \pm 1$.  Similarly, the condition
$\Re({\bf A}_3^*\cdot {\bf A}_1)=0$ implies that $\beta_2 = \pi +
\sigma_2 \beta_1$, where $\sigma_2 = \pm 1$.  If $\sigma_1 =
\sigma_2$, the condition $\Re({\bf A}_2^*\cdot {\bf A}_3)=0$ becomes
$\cos( \alpha_1-\beta_1)=0$. Referring to Eq.~(\ref{eqn:densities}),
we observe that if $n_1$ is to have four-fold rotational symmetry,
$\cos \alpha_1$ and $\cos\beta_1$ must vanish as well. This is not
possible.  If $\sigma_1 = - \sigma_2$, on the other hand, the
condition $\Re({\bf A}_2^*\cdot {\bf A}_3)=0$ becomes $\cos \alpha_1
\cos\beta_1 = 0$. Equation~(\ref{eqn:densities}) shows that if $n_1$
is to have four-fold rotational symmetry, it is required that $\cos
\alpha_1 = \cos\beta_1 = \cos( \alpha_1-\beta_1) = 0$, which is an
impossibility.  We conclude that there is no solution with both
four-fold rotational and component symmetries for two components.


\begin{thebibliography}{10}

\bibitem{leggett2001}
A.~J. Leggett, Rev. Mod. Phys. {\bf 73},  307  (2001).

\bibitem{lewenstein2006}
M. Lewenstein, A. Sanpera, V. Ahufinger, B. Damski, A. Sen, and U. Sen,
  Advances in Physics {\bf 56},  243  (2007).

\bibitem{matthews1999}
M.~R. Matthews, B.~P. Anderson, P.~C. Haljan, D.~S. Hall, C.~E. Wieman, and
  E.~A. Cornell, Phys. Rev. Lett. {\bf 83},  2498  (1999).

\bibitem{williams1999}
J.~E. Williams and M.~J. Holland, Nature {\bf 401},  568  (1999).

\bibitem{higbie2005}
J.~M. Higbie, L.~E. Sadler, S. Inouye, A.~P. Chikkatur, S.~R. Leslie, K.~L.
  Moore, V. Savalli, and D.~M. Stamper-Kurn, Phys. Rev. Lett. {\bf 95},  050401
   (2005).

\bibitem{widera2005}
A. Widera, F. Gerbier, S. Folling, T. Gericke, O. Mandel, and I. Bloch, Phys.
  Rev. Lett. {\bf 95},  190405  (2005).

\bibitem{carr2001b}
J.~C. Bronski, L.~D. Carr, B. Deconinck, and J.~N. Kutz, Phys. Rev. Lett. {\bf
  86},  1402  (2001).

\bibitem{carr2001c}
J.~C. Bronski, L.~D. Carr, B. Deconinck, J.~N. Kutz, and K. Promislow, Phys.
  Rev. E {\bf 63},  036612  (2001).

\bibitem{carr2001d}
J.~C. Bronski, L.~D. Carr, R. Carretero-Gonz\'alez, B. Deconinck, J.~N. Kutz,
  and K. Promislow, Phys. Rev. E {\bf 64},  056615  (2001).

\bibitem{deconinck2001}
B. Deconinck, B.~A. Frigyik, and J.~N. Kutz, Phys. Lett. A {\bf 283},  177
  (2001).

\bibitem{deconinck2002}
B. Deconinck, B.~A. Frigyik, and J.~N. Kutz, J. Nonlinear Sci. {\bf 12},  169
  (2002).

\bibitem{hai2004}
W. Hai, C. Lee, X. Fang, and K. Gao, Physica A {\bf 335},  445  (2004).

\bibitem{deconinck2003}
B. Deconinck, J.~N. Kutz, M.~S. Patterson, and B.~W. Warner, J. Phys. A: Math.
  Gen. {\bf 36},  5431  (2003).

\bibitem{bradleyRM2005}
R.~M. Bradley, B. Deconinck, and J.~N. Kutz, J. Phys. A: Math. Gen. {\bf 38},
  1901  (2005).

\bibitem{dalfovo1999}
F. Dalfovo, S. Giorgini, L.~P. Pitaevskii, and S. Stringari, Rev. Mod. Phys.
  {\bf 71},  463  (1999).

\bibitem{mur-petit2006}
J. Mur-Petit, M. Guilleumas, A. Polls, A. Sanpera, M. Lewenstein, K. Bongs, and
  K. Sengstock, Phys. Rev. A {\bf 73},  013629  (2006).

\bibitem{footnote1}
See Ref.~\cite{rey2004} for a detailed derivation of $t_h$ and $U$ from first
  principles quantum field theory.

\bibitem{footnote3}
Note that $t_h$ is normally called $t$ in the condensed matter literature and
  is sometimes denoted $J$ in the case of ultracold quantum gases. The $J$
  notation leads to a ``$J-J$'' model instead of a $t-J$ model, and so we avoid
  it. We will reserve $t$ for time as is standard in dynamics, and use $t_h$
  for the hopping energy.

\bibitem{rey2004}
A.~M. Rey, Ph.D. thesis, University of Maryland, 2004.

\bibitem{carr2007a}
R.~V. Mishmash and L.~D. Carr, Phys. Rev. Lett.  , under review; e-print
  http://arxiv.org/abs/0710.0045 (2007).

\bibitem{olshanii1998}
M. Olshanii, Phys. Rev. Lett. {\bf 81},  938  (1998).

\bibitem{carr2000e}
L.~D. Carr, M.~A. Leung, and W.~P. Reinhardt, J. Phys. B: At. Mol. Opt. Phys.
  {\bf 33},  3983  (2000).

\bibitem{petrov2000}
D.~S. Petrov, M. Holzmann, and G.~V. Shlyapnikov, Phys. Rev. Lett. {\bf 84},
  2551  (2000).

\bibitem{petrov2000b}
D.~S. Petrov, G.~V. Shlyapnikov, and J.~T.~M. Walraven, Phys. Rev. Lett. {\bf
  85},  3745  (2000).

\bibitem{HoTL1998}
T.-L. Ho, Phys. Rev. Lett. {\bf 81},  742  (1998).

\bibitem{ohmi1998}
T. Ohmi and K. Machida, J. Phys. Soc. Japan {\bf 67},  1822  (1998).

\bibitem{law1998}
C.~K. Law, H. Pu, and N.~P. Bigelow, Phys. Rev. Lett. {\bf 81},  5257  (1998).

\bibitem{grimm2000}
R. Grimm, M. Weidemuller, and Y.~B. Ovchinnikov, Adv. Atom. Mol. Opt. Phys.
  {\bf 42},  95  (2000).

\bibitem{hemmerlich1992}
A. Hemmerlich, D. Schropp, T. Esslinger, and T.~W. H\"ansch, Europhys. Lett.
  {\bf 18},  391  (1992).

\bibitem{hall1998}
D.~S. Hall, M.~R. Matthews, J.~R. Ensher, C.~E. Wieman, and E.~A. Cornell,
  Phys. Rev. Lett. {\bf 81},  1539  (1998).

\bibitem{manakov1974}
S.~V. Manakov, Sov. Phys. JETP {\bf 38},  693  (1974).

\bibitem{roberts}
J.~N. Roberts, private communication.

\bibitem{myatt1997}
C.~J. Myatt, E.~A. Burt, R.~W. Ghrist, E.~A. Cornell, and C.~E. Wieman, Phys.
  Rev. Lett. {\bf 78},  586  (1997).

\bibitem{stamper1998}
D.~M. Stamper-Kurn, M.~R. Andrews, A.~P. Chikkatur, S. Inouye, H.-J. Miesner,
  J. Stenger, and W. Ketterle, Phys. Rev. Lett. {\bf 80},  2027  (1998).

\bibitem{stenger1998}
J. Stenger, S. Inouye, D.~M. Stamper-Kurn, H.-J. Miesner, A.~P. Chikkatur, and
  W. Ketterle, Nature {\bf 396},  345  (1998).

\bibitem{burke1998}
J.~P. Burke, Jr., C.~H. Greene, and J.~L. Bohn, Phys. Rev. Lett. {\bf 81},
  3355  (1998).

\bibitem{footnote2}
Note, however, that it is a simple matter to write down the corresponding
  $\sigma = -1$ solutions.

\bibitem{deconinck2004}
B. Deconinck, P.~G. Kevrekidis, H.~E. Nistazakis, and D.~J. Frantzeskakis,
  Phys. Rev. A {\bf 70},  063605  (2004).

\bibitem{carr2005d}
B.~T. Seaman, L.~D. Carr, and M.~J. Holland, Phys. Rev. A {\bf 71},  033622
  (2005).

\bibitem{carr2005c}
L.~D. Carr, M.~J. Holland, and B.~A. Malomed, J. Phys. B: At. Mol. Opt. {\bf
  38},  3217  (2005).

\bibitem{footnote4}
We use the driver routine \textsc{odeint} and the stepper routine \textsc{rkqs}
  of \cite{press1993}.

\bibitem{press1993}
W.~H. Press, S.~A. Teukolsky, W.~T. Vetterling, and B.~P. Flannery, {\em
  Numerical Recipes in C: The Art of Scientific Computing} (Cambridge Univ.
  Press, Cambridge, U.K., 1993).

\bibitem{carr2005e}
B.~T. Seaman, L.~D. Carr, and M.~J. Holland, Phys. Rev. A {\bf 72},  033602
  (2005).

\bibitem{footnote5}
For very weak potentials, with $\mathcal V/E_R$ below about $0.8$,
  corresponding to particle densities less than about $260$ per well,
  instability-onset times on the boundary of the inaccessible region exceeded
  the threshold defining the boundary of the stable region. However, that
  threshold was not reached for $500$ or $1000$ particles per well with
  potentials as weak as $\mathcal V/E_R=0.16$. Evidently the structure of the
  stability map is more complicated in the far lower-left corner of
  \protect\figref{fig:instab_2s1d_nv} than in the rest of the plane.

\bibitem{krugerP2007}
P. Kr\"uger, Z. Hadzibabic, and J. Dalibard, Phys. Rev. Lett. {\bf 99},  040402
   (2007).

\bibitem{sebbystrabley2006}
J. Sebby-Strabley, M. Anderlini, P.~S. Jessen, and J.~V. Porto, Phys. Rev. A
  {\bf 73},  033605  (2006).

\end{thebibliography}
\end{document}